\documentstyle[12pt,epsf]{article}

\renewcommand{\bar}{\overline}

\def\ru1{\rule[-0.4truecm]{0mm}{1truecm}}

\renewcommand{\bar}[1]{\overline{#1}}
\newcommand{\M}{{\cal M}}
\newcommand{\VEV}[1]{\left\langle{#1}\right\rangle}
\newcommand{\etal}{{\em et al.}}
\newcommand{\ie}{{\it i.e.}}
\newcommand{\eg}{{\it e.g.}}
\newcommand{\half}{{1\over 2}}

\newcommand{\lsim} {\buildrel < \over {_\sim}}
\newcommand{\ket}[1]{\vert\,{#1}\rangle}

\renewcommand{\bar}{\overline}

\newcommand{\hsim}[1]{\ \ \mathrel{\rlap{\lower-4pt\hbox{$\sim$}}
                    \hskip-3pt\hbox{$#1$}}\,}

\newcommand{\eqx} {\buildrel = \over {_{x \to 1}}}
\newcommand{\gsim} {\buildrel > \over {_\sim}}

\begin {document}
\begin{flushright}
{\small
SLAC--PUB--8627\\
September 2000\\}
\end{flushright}

\begin{center}
{{\bf\LARGE   
The Light-Cone Fock Expansion in Quantum Chromodynamics}
\footnote{Work supported by Department of Energy contract  DE--AC03--76SF00515.}}

\bigskip
{\it Stanley J. Brodsky \\
Stanford Linear Accelerator Center \\
Stanford University, Stanford, California 94309 \\
E-mail:  sjbth@slac.stanford.edu}
\medskip
\end{center}

\vfill

\begin{center}
{\bf\large   
Abstract }
\end{center}

A fundamental question in QCD is the non-perturbative structure of hadrons
at the amplitude level---not just the single-particle flavor,
momentum, and helicity distributions of the quark constituents,  but also
the multi-quark, gluonic, and hidden-color correlations intrinsic to
hadronic and nuclear wavefunctions.  The light-cone Fock-state
representation of QCD encodes the properties of a hadrons in terms of
frame-independent wavefunctions.  A number of applications are
discussed, including semileptonic $B$ decays, deeply virtual Compton
scattering, and dynamical higher twist effects in inclusive reactions.  A
new type of jet production reaction, ``self-resolving diffractive
interactions" can provide direct information on the light-cone
wavefunctions of hadrons in terms of their quark and gluon degrees of
freedom as well as the composition of nuclei in terms of their nucleon
and mesonic degrees of freedom.  The relation of the intrinsic sea to the
light-cone wavefunctions is discussed.  The physics of light-cone
wavefunctions is illustrated for the quantum fluctuations of an electron.

\vfill

\begin{center} 
{\it Presented at  \\
VII HADRON PHYSICS 2000  \\
 Caraguatatuba, S${\tilde a}$o Paulo, Brazil\\
April 10--15, 2000}\\
\end{center}

\vfill

\newpage

\section{Introduction}

Quantum Chromodynamics, the non-abelian $SU(N_C = 3)$ gauge theory of
quark and gluons is the central theory of particle and nuclear physics.
The range of applications of QCD to physical processes is extraordinary,
ranging from the dynamics and structure of hadrons and nuclei, the
properties of electroweak transitions, quark and gluon jet processes,  to
the properties and phases of hadronic matter at the earliest stages of the
universe.  At very short distances QCD is believed to unify with the
electroweak interactions, and possibly even gravity, into more
fundamental theories.

There has been enormous progress in understanding QCD since its
inception in 1973,\cite{Fritzsch:1973pi} particularly in the applications of the
perturbative theory to inclusive and exclusive processes
involving collisions at large momentum transfer.  New
experimental tools are continually being developed which probe the
non-perturbative structure of the theory, such as hard
diffractive reactions, self-resolving jet reactions,
semi-exclusive reactions, deeply virtual Compton scattering, and
heavy ion collisions.  Nevertheless, many fundamental
questions have not been resolved.  These include
rigorous proofs of color confinement, the behavior of the QCD coupling
at small scales, the computation of the non-perturbative structure of
hadrons in terms of their quark and gluon degrees of freedom,  the
problem of $n!$ growth of the perturbation theory (renormalon phenomena), the
nature of the pomeron and reggeons, the nature of shadowing
and antishadowing in nuclear collisions, the apparent conflict between QCD
vacuum structure and the small size of the cosmological constant, and the
problems of scale and scheme ambiguities in perturbative QCD expansion.
One of the most pressing problems is to understand the QCD physics of
exclusive $B$-meson decays at the amplitude level, since the interpretation
of the basic parameters of the electroweak theory and $CP$ violation
depend on hadronic  dynamics and phase structure.

The most challenging nonperturbative problem
in QCD is the solution of the bound state problem;
\ie, to determine the structure and spectrum of hadrons and nuclei in
terms of their quark and gluon degrees of freedom.  Ideally, one wants a
frame-independent, quantum-mechanical description of had\-rons at the
amplitude level capable of encoding multi-quark, hidden-color and gluon
momentum, helicity, and flavor correlations in the form of universal
process-independent hadron wavefunctions.  Remarkably, the light-cone Fock
expansion allows just such a unifying representation.

Formally, the light-cone expansion is constructed by quantizing QCD at
fixed light-cone time \cite{Dirac:1949cp} $\tau = t + z/c$ and forming the
invariant light-cone
Hamiltonian: $ H^{QCD}_{LC} = P^+ P^- - {\vec P_\perp}^2$ where
$P^\pm = P^0 \pm P^z$.\cite{PinskyPauli}   The operator
$P^- = i {d\over d\tau}$ generates light-cone time translations.
The
$P^+$ and
$\vec P_\perp$ momentum operators are independent of
the interactions.  Each intermediate state consists of particles with
light-cone energy $k^- = {\vec k^2_\perp + m^2\over k^+} > 0$ and positive
$k^+$.

The  procedure for quantizing non-Abelian gauge theory in QCD is
well-known.\cite{BL80,bro}  In brief: if one chooses light-cone gauge
$A^{+}=0$, the dependent gauge field $A^{-}$ and quark field $\psi^{-} =
\Lambda^{-} \psi$ can be eliminated in terms of the physical transverse
field $A^{\perp}$ and
$A^{+} =
\Lambda^{+} \psi$ fields.  Here
$\,\Lambda^{\pm}= {1\over 2} \gamma^{\mp}\gamma^{\pm} \, $ are hermitian
projection operators.  Remarkably, no ghosts fields appear in the
formalism, since only physical degrees of freedom propagate.  The
interaction Hamiltonian includes the usual Dirac interactions between the
quarks and gluons, the three-point and four-point gluon non-Abelian
interactions plus instantaneous \cite{bro} light-cone time gluon exchange
and quark exchange contributions:
\begin{eqnarray}
{\cal H}_{int}&=&
  -g \,{{\bar\psi}}^{i}
\gamma^{\mu}{A_{\mu}}^{ij}{{\psi}}^{j}   \nonumber \\
&& +\frac{g}{2}\,
f^{abc} \,(\partial_{\mu}{A^{a}}_{\nu}-
\partial_{\nu}{A^{a}}_{\mu}) A^{b\mu} A^{c\nu} \nonumber \\
&& +\frac {g^2}{4}\,
f^{abc}f^{ade} {A_{b\mu}} {A^{d\mu}} A_{c\nu} A^{e\nu} \nonumber \\
&& - \frac{g^{2}}{ 2}\,\, {{\bar\psi}}^{i}
\gamma^{+}
\,(\gamma^{\perp'}{A_{\perp'}})^{ij}\,\frac{1}{i\partial_{-}} \,
(\gamma^{\perp} {A_{\perp}})^{jk}\,{\psi}^{k} \nonumber \\
&& -\frac{g^{2}}{ 2}\,{j^{+}}_{a}\, \frac {1}{(\partial_{-})^{2}}\,
{j^{+}}_{a}
\end{eqnarray}
where
\begin{equation}
{j^{+}}_{a}={{\bar\psi}}^{i}
\gamma^{+} ( {t_{a}})^{ij}{{\psi}}^{j}
+ f_{abc} (\partial_{-}
A_{b\mu}) A^{c\mu}
\end{equation}
Srivastava and I have recently shown how one can use the Dyson-Wick
formalism to construct the Feynman rules in light-cone gauge for QCD.
The gauge fields satisfy both the light-cone gauge and the Lorentz
condition $\partial_\mu A^\mu =0.$  We have also shown that one can also
effectively quantize QCD in the covariant Feynman gauge.\cite{Srivastava:2000gi}

The eigen-spectrum of $ H^{QCD}_{LC}$ in principle gives the entire mass
squared spectrum of color-singlet hadron states in QCD, together with
their respective light-cone wavefunctions.  For example, the
proton state satisfies:
$ H^{QCD}_{LC} \ket{\Psi_p} = M^2_p \ket{\Psi_p}$.
The projection of
the proton's eigensolution $\ket{\Psi_p}$ on the color-singlet
$B = 1$, $Q = 1$ eigenstates $\{\ket{n} \}$
of the free Hamiltonian $ H^{QCD}_{LC}(g = 0)$ gives the
light-cone Fock expansion: \cite {BrodskyLepage}
\begin{eqnarray}
\left\vert \Psi_p; P^+, {\vec P_\perp}, \lambda \right> &=&
\sum_{n \ge 3,\lambda_i}  \int \Pi^{n}_{i=1}
{d^2k_{\perp i} dx_i \over \sqrt{x_i} 16 \pi^3} \cr
&& 16 \pi^3 \delta\left(1- \sum^n_j x_j\right) \delta^{(2)}
\left(\sum^n_\ell \vec k_{\perp \ell}\right) \cr
&&\left\vert n;
x_i P^+, x_i {\vec P_\perp} + {\vec k_{\perp i}}, \lambda_i\right >
\psi_{n/p}(x_i,{\vec k_{\perp i}},\lambda_i)
 . \nonumber
\end{eqnarray}
The light-cone Fock
wavefunctions $\psi_{n/H}(x_i,{\vec k_{\perp i}},\lambda_i)$
thus interpolate between the hadron $H$ and
its quark and gluon degrees of freedom.
The light-cone momentum fractions of the constituents,
$x_i = k^+_i/P^+$ with $\sum^n_{i=1} x_i = 1,$ and the transverse
momenta ${\vec k_{\perp i}}$ with
$\sum^n_{i=1} {\vec k_{\perp i}} = {\vec 0_\perp}$ appear as
the momentum
coordinates of the light-cone Fock wavefunctions.  A crucial feature is
the frame-independence of the light-cone wavefunctions.  The $x_i$ and
$\vec k_{\perp i}$ are relative coordinates independent of the hadron's
momentum $P^\mu$.  The actual physical transverse momenta are
${\vec p_{\perp i}} = x_i {\vec P_\perp} + {\vec k_{\perp i}}.$

The $\lambda_i$ label the light-cone spin $S^z$ projections of the quarks and
gluons along the $z$ direction.  The physical gluon
polarization vectors
$\epsilon^\mu(k,\ \lambda = \pm 1)$ are specified in light-cone
gauge by the conditions $k \cdot \epsilon = 0,\ \eta \cdot \epsilon =
\epsilon^+ = 0.$
Each light-cone Fock wavefunction satisfies conservation of the
$z$ projection of angular momentum:
$
J^z = \sum^n_{i=1} S^z_i + \sum^{n-1}_{j=1} l^z_j \ .
$
The sum over $S^z_i$
represents the contribution of the intrinsic spins of the $n$ Fock state
constituents.  The sum over orbital angular momenta
$l^z_j = -{\mathrm i} (k^1_j\frac{\partial}{\partial k^2_j}
-k^2_j\frac{\partial}{\partial k^1_j})$ derives from
the $n-1$ relative momenta.  This excludes the contribution to the
orbital angular momentum due to the motion of the center of mass, which
is not an intrinsic property of the hadron.\cite{Brodsky:2000ii}

Light-cone wavefunctions are thus the frame-independent interpolating
functions between hadron and quark and gluon degrees of freedom.  Hadron
amplitudes are computed from the convolution of the light-cone
wavefunctions with irreducible quark-gluon amplitudes.  For example,
space-like form factors can be represented as the diagonal $\Delta n = 0$
overlap of light-cone wavefunctions.  Time-like form factors such as
semi-exclusive $B$ decays can be expressed as the sum of diagonal $\Delta
n = 0$ and $\Delta n = 2$ overlap integrals.  Structure functions are
simply related to the sum over absolute squares of the light-cone
wavefunctions.  More generally, all multi-quark and gluon correlations in
the bound state are represented by the light-cone wavefunctions.  Thus in
principle, all of the complexity of a hadron is encoded in the light-cone
Fock representation, and  the light-cone Fock representation is thus a
representation of the underlying quantum field theory.

The LC wavefunctions $\psi_{n/H}(x_i,\vec
k_{\perp i},\lambda_i)$ are universal, process-independent, and thus
control all hadronic reactions.  In the case of deep inelastic scattering,
one needs to evaluate the imaginary part of the virtual Compton
amplitude ${\cal M}[\gamma^*(q) p \to \gamma^* (q) p].$ The simplest
frame choice for electroproduction is
$q^+ = 0, q_\perp^2 = Q^2= - q^2,  q^- = {2 q\cdot p / P^+},
p^+ = P^+, p_\perp = 0_\perp, p^- = {M_p^2/ P^+}.  $ At leading twist,
soft final-state interactions of the outgoing hard quark line are
power-law suppressed in light-cone gauge, so the calculation of the
virtual Compton amplitude reduces to the evaluation of matrix elements of
the products of free quark currents of the free quarks.  The absorptive
amplitude imposes conservation of light-cone energy:
$p^- +  q^- = \sum^n_i k^-_i$ for the $n-$particle Fock state.  In the
impulse approximation, where only one quark $q$ recoils against the
scattered lepton, this condition becomes
\begin{equation}
M_p^2 + 2 q\cdot p = {({\vec k}_{\perp q}+ {\vec q}_\perp)^2 + m_q^2
\over x_q} +
\sum_{i \ne q} {k_{\perp i}^2 + m_i^2 \over x_i} \ .
\end{equation}
If we neglect the transverse momenta $k^2_\perp$ relative to $Q^2$ in the
Bjorken limit $Q^2 \to \infty,$
$x_{bj} = {Q^2/ 2 q \cdot p}$ fixed,  we obtain the condition $x_q =
x_{bj}$; {\it i.e.}, the light-cone fraction $x_q= k^+/p^+$
of the struck quark is
kinematically fixed to be equal to the Bjorken ratio.
Contributions from high
$k^2_\perp = {\cal O}(Q^2)$ which originate from the perturbative QCD
radiative corrections to the struck quark line lead to the DGLAP
evolution equations.

Thus given the light-cone wavefunctions, one can compute\cite{BL80}
all of the leading twist helicity and
transversity distributions measured in polarized deep inelastic
lepton scattering.\cite{Jaffe:1989up}  For example,
the helicity-specific quark distributions at resolution $\Lambda$
correspond to
\begin{eqnarray}
q_{\lambda_q/\Lambda_p}(x, \Lambda)
&=& \sum_{n,q_a}
\int\prod^n_{j=1} {dx_j d^2 k_{\perp j}\over 16 \pi^3} \sum_{\lambda_i}
\vert \psi^{(\Lambda)}_{n/H}(x_i,\vec k_{\perp i},\lambda_i)\vert^2
\\
&& \times 16 \pi^3 \delta\left(1- \sum^n_i x_i\right) \delta^{(2)}
\left(\sum^n_i \vec k_{\perp i}\right)
\delta(x - x_q) \delta_{\lambda, \lambda_q}
\Theta(\Lambda^2 - {\cal M}^2_n)\ , \nonumber
\end{eqnarray}
where the sum is over all quarks $q_a$ which match the quantum
numbers, light-cone momentum fraction $x,$ and helicity of the struck
quark.  Similarly, the transversity distributions and
off-diagonal helicity convolutions are defined as a density matrix of the
light-cone wavefunctions.  This defines the LC
factorization scheme \cite{BL80} where the
invariant mass squared ${\cal M}^2_n = \sum_{i = 1}^n {(k_{\perp i}^2 +
m_i^2 )/ x_i}$ of the $n$ partons of the light-cone wavefunctions are
limited to $ {\cal M}^2_n < \Lambda^2$

The light-cone wavefunctions also specify the multi-quark and gluon
correlations of the hadron.  For example,  the distribution of spectator
particles in the final state which could be measured in the proton
fragmentation region in deep inelastic scattering at an electron-proton
collider are in principle encoded in the light-cone wavefunctions.
We also note that the high momentum tail of the light-cone wavefunctions
can be computed perturbatively in QCD.  In particular, the
evolution equations for structure functions and distribution amplitudes
follow from the perturbative high transverse momentum behavior of the
light-cone wavefunctions.\cite{BrodskyLepage} The gauge theory features
of color transparency and color opacity for color singlet hadrons follows
from the distribution of the quarks and gluons in transverse space of the
hadron wavefunctions.\cite{BM}

There are many sources of power-law corrections to the standard
leading twist formula for deep inelastic structure functions.
Higher-twist corrections arise from QCD radiative corrections
(renormalons), final-state interactions, finite target mass effects
\cite{Nachtmann:1973mr}, constituent masses, and their transverse
momenta $k_\perp.$\cite{Brodsky:1979gy}  Despite the many sources
of power-law corrections to the deep inelastic cross section, certain
types of dynamical contributions will stand out at large $x_{bj}$ since
they arise from compact, highly-correlated fluctuations of the proton
wavefunction.  In particular, as I will discuss in Section 12, there are
particularly interesting dynamical ${\cal O}(1/Q^2)$ corrections which
are due to the {\it interference} of quark currents; {\it i.e.},
contributions which involve
leptons scattering amplitudes from two different quarks.

Recently, the E791 experiment at Fermilab has demonstrated that the
light-cone wavefunction of a hadron can be directly measured by
diffractively dissociating a high energy hadron into jets.\cite{Ashery:1999nq}
I will review the physics of self-resolving
interactions in Section 12.

In addition to the light-cone Fock expansion, a number of other
useful theoretical tools are available to eliminate theoretical
ambiguities in QCD predictions:

(1) Conformal symmetry provides a template for
QCD predictions,\cite{Brodsky:1999gm} leading to relations between
observables which are present even in a theory which is not scale
invariant.  For example, the natural representation of distribution
amplitudes is in terms of an expansion of orthonormal conformal functions
multiplied by anomalous dimensions determined by QCD evolution
equations.\cite{Brodsky:1980ny,Muller:1994hg,Braun:1999te} Thus an
important guide in QCD analyses is to identify the underlying conformal
relations of QCD which are manifest if we drop quark masses and effects
due to the running of the QCD couplings.  In fact, if QCD has an infrared
fixed point (vanishing of the Gell Mann-Low function at low momenta), the
theory will closely resemble a scale-free conformally symmetric theory in
many applications.

(2) Commensurate scale relations\cite{Brodsky:1995eh,Brodsky:1998ua} are
perturbative QCD predictions which relate observable to observable at
fixed relative scale, such as the ``generalized Crewther relation"
\cite{Brodsky:1996tb}, which connects the Bjorken and Gross-Llewellyn
Smith deep inelastic scattering sum rules to measurements of the $e^+
e^-$ annihilation cross section.  The relations have no renormalization
scale or scheme ambiguity.  The coefficients in the perturbative series
for commensurate scale relations are identical to those of conformal QCD;
thus no infrared renormalons are present.\cite{Brodsky:1999gm}  One can
identify the required conformal coefficients at any finite order by
expanding the coefficients of the usual PQCD expansion around a formal
infrared fixed point, as in the Banks-Zak method.\cite{Brodsky:2000cr}
All non-conformal effects are absorbed by fixing the ratio of the
respective momentum transfer and energy scales.  In the case of
fixed-point theories, commensurate scale relations relate both the ratio
of couplings and the ratio of scales as the fixed point is approached.
\cite{Brodsky:1999gm}

(3) $\alpha_V$ and Skeleton Schemes.  A physically natural scheme for
defining the QCD coupling in exclusive and other processes is the
$\alpha_V(Q^2)$ scheme defined from the potential of static heavy
quarks.  Heavy-quark lattice gauge theory can provide highly precise
values for the coupling.  All vacuum polarization corrections due to
fermion pairs are then automatically and analytically incorporated into
the Gell Mann-Low function, thus avoiding the problem of explicitly
computing and resumming quark mass corrections related to the running of
the coupling.\cite{bgmr} The use of a finite effective charge such as
$\alpha_V$ as the expansion parameter also provides a basis for
regulating the infrared nonperturbative domain of the QCD coupling.  A
similar coupling and scheme can be based on an assumed skeleton expansion
of the theory.\cite{Brodsky:2000cr}

(4) The Abelian Correspondence Principle.  One can consider QCD
predictions as analytic functions of the number of colors $N_C$ and
flavors $N_F$.  In particular, one can show at all orders of
perturbation theory that PQCD predictions reduce to those of an Abelian
theory at $N_C \to 0$ with ${\widehat \alpha} = C_F \alpha_s$ and
${\widehat N_F} = N_F/T C_F$ held fixed.\cite{Brodsky:1997jk}  There is
thus a deep connection between QCD processes and their corresponding QED
analogs.

A review of these topics can be found in the lectures by Rathsman
and myself.  \cite{Brodsky:1999bd,Brodsky:1999gm}

\section{Applications of Light-cone wavefunctions to 
\hfill\break Current Matrix Elements}

As I shall review in the next sections, the light-cone Fock representation
of current matrix elements has a number of simplifying properties.  Matrix
elements of space-like local operators for the coupling of photons,
gravitons and the deep inelastic structure functions can all be expressed
as overlaps of light-cone wavefunctions with the same number of Fock
constituents.  This is possible since one can choose the special frame
$q^+ = 0$
\cite{Drell:1970km,West:1970av} for space-like momentum transfer and
take matrix elements of ``plus" components of currents such as $J^+$ and
$T^{++}$.  Since the physical vacuum in light-cone quantization coincides
with the perturbative vacuum, no contributions to matrix elements from
vacuum fluctuations occur.\cite{Brodsky:1998de}   Exclusive
semi-leptonic
$B$-decay amplitudes involving time-like currents such as $B\rightarrow A
\ell
\bar{\nu}$ can also be evaluated exactly.\cite{Brodsky:1999hn,Ji:1999gt}
In this case, the time-like decay matrix elements require the
computation of both the diagonal matrix element $n \rightarrow n$ where
parton number is conserved and the off-diagonal $n+1\rightarrow n-1$
convolution such that the current operator annihilates a $q{\bar{q'}}$
pair in the initial $B$ wavefunction.  This term is a consequence of the
fact that the time-like decay $q^2 = (p_\ell + p_{\bar{\nu}} )^2 > 0$
requires a positive light-cone momentum fraction $q^+ > 0$.  A similar
result holds for the light-cone wavefunction representation of the deeply
virtual Compton amplitude.\cite{DVCSHwang}

\section{Electromagnetic and Gravitational Form Factors}

The light-cone Fock representation allows one to compute all matrix
elements of local currents as overlap integrals of the light-cone Fock
wavefunctions.  In particular, we can evaluate forward and
non-forward matrix elements of the electroweak currents,  moments of the
deep inelastic structure functions,  as well as the electromagnetic form
factors and the magnetic moment.  Given the local operators for the
energy-momentum tensor
$T^{\mu \nu}(x)$ and the angular momentum tensor
$M^{\mu \nu \lambda}(x)$, one can directly compute
momentum fractions, spin properties, the gravitomagnetic
moment, and the form factors $A(q^2)$ and $B(q^2)$ appearing in the
coupling of gravitons to composite systems.

In the case of a spin-${1\over 2}$  composite system, the Dirac and
Pauli form factors $F_1(q^2)$ and $F_2(q^2)$ are defined by
\begin{equation}
      \langle P'| J^\mu (0) |P\rangle
       = \bar u(P')\, \Big[\, F_1(q^2)\gamma^\mu +
F_2(q^2){i\over 2M}\sigma^{\mu\alpha}q_\alpha\, \Big] \, u(P)\ ,
\label{Drell1}
\end{equation}
where $q^\mu = (P' -P)^\mu$ and $u(P)$ is the bound state spinor.
In the light-cone formalism it is convenient to identify the Dirac and
Pauli form factors from the
helicity-conserving and helicity-flip vector current matrix elements of
the $J^+$ current
\cite{BD80}:
\begin{equation}
\VEV{P+q,\uparrow\left|\frac{J^+(0)}{2P^+}
\right|P,\uparrow} =F_1(q^2) \ ,
\label{BD1}
\end{equation}
\begin{equation}
\VEV{P+q,\uparrow\left|\frac{J^+(0)}{2P^+}\right|P,\downarrow}
=-(q^1-{\mathrm i} q^2){F_2(q^2)\over 2M}\ .
\label{BD2}
\end{equation}
The magnetic moment of a composite system is one of its
most basic properties.  The magnetic moment is defined at the $q^2 \to 0$
limit,
\begin{equation}
\mu=\frac{e}{2 M}\left[ F_1(0)+F_2(0) \right] ,
\label{DPmu}
\end{equation}
where $e$ is the charge and $M$ is the mass of the composite
system.  We use the standard light-cone frame
($q^{\pm}=q^0\pm q^3$):
\begin{eqnarray}
q &=& (q^+,q^-,{\vec q}_{\perp}) = \left(0, \frac{-q^2}{P^+},
{\vec q}_{\perp}\right), \nonumber \\
P &=& (P^+,P^-,{\vec P}_{\perp}) = \left(P^+, \frac{M^2}{P^+},
{\vec 0}_{\perp}\right),
\label{LCF}
\end{eqnarray}
where $q^2=-2 P \cdot q= -{\vec q}_{\perp}^2$ is 4-momentum square
transferred by the photon.

The Pauli form factor and the anomalous magnetic moment $\kappa =
{e\over 2 M} F_2(0)$ can then be  calculated from the
expression
\begin{equation}
-(q^1-{\mathrm i} q^2){F_2(q^2)\over 2M} =
\sum_a  \int
{{\mathrm d}^2 {\vec k}_{\perp} {\mathrm d} x \over 16 \pi^3}
\sum_j e_j \ \psi^{\uparrow *}_{a}(x_i,{\vec k}^\prime_{\perp
i},\lambda_i) \,
\psi^\downarrow_{a} (x_i, {\vec k}_{\perp i},\lambda_i)
{}\ ,
\label{LCmu}
\end{equation}
where  the summation is over all contributing Fock states $a$ and struck
constituent charges $e_j$.  The arguments of the final-state
light-cone  wavefunction are
\cite{DY70,West:1970av}
\begin{equation}
{\vec k}'_{\perp i}={\vec k}_{\perp i}+(1-x_i){\vec
q}_{\perp}
\label{kprime1}
\end{equation}
for the struck constituent and
\begin{equation}
{\vec k}'_{\perp i}={\vec k}_{\perp i}-x_i{\vec q}_{\perp}
\label{kprime2}
\end{equation}
for each spectator.
Notice that the magnetic moment must be calculated from the
spin-flip non-forward matrix element of the current.  It is not given by a
diagonal forward matrix element.\cite{Che98}
In the ultra-relativistic limit where the radius of the system is small
compared to its Compton scale $1/M$, the anomalous magnetic moment must
vanish.\cite{Bro94}  The light-cone formalism is consistent with this
theorem.

The form factors of the energy-momentum tensor for a spin-$\half$ \
composite are defined by
\begin{eqnarray}
      \langle P'| T^{\mu\nu} (0)|P \rangle
       &=& \bar u(P')\, \Big[\, A(q^2)
       \gamma^{(\mu} \bar P^{\nu)} +
   B(q^2){i\over 2M} \bar P^{(\mu} \sigma^{\nu)\alpha}
q_\alpha \nonumber \\
   &&\qquad\qquad +  C(q^2){1\over M}(q^\mu q^\nu - g^{\mu\nu}q^2)
    \, \Big]\, u(P) \ ,
\label{Ji12}
\end{eqnarray}
where $q^\mu = (P'-P)^\mu$,
$\bar P^\mu={1\over 2}(P'+P)^\mu$,
$a^{(\mu}b^{\nu)}={1\over 2}(a^\mu b^\nu +a^\nu b^\mu)$,
and $u(P)$ is the spinor of the system.

As in the
light-cone decomposition  Eqs.  (\ref{BD1}) and (\ref{BD2})
of the Dirac and Pauli form factors for the vector
current \cite{BD80}, we can obtain  the light-cone representation
of the $A(q^2)$ and $B(q^2)$ form factors of the energy-tensor
Eq. (\ref{Ji12}).
Since we  work in the interaction picture, only the non-interacting
parts of the energy momentum tensor $T^{+ +}(0)$ need to be computed in
the light-cone formalism.
By calculating the $++$ component of
Eq. (\ref{Ji12}), we find
\begin{equation}
\VEV{P+q,\uparrow\left|\frac{T^{++}(0)}{2(P^+)^2}
\right|P,\uparrow} =A(q^2)\ ,
\label{eBD1}
\end{equation}
\begin{equation}
\VEV{P+q,\uparrow\left|\frac{T^{++}(0)}{2(P^+)^2}\right|P,\downarrow}=
-(q^1-{\mathrm i} q^2){B(q^2)\over 2M}\ .
\label{eBD2}
\end{equation}
The $A(q^2)$ and $B(q^2)$ form factors
Eqs. (\ref{eBD1}) and (\ref{eBD2})
are similar to the $F_1(q^2)$ and $F_2(q^2)$ form
factors Eqs.  (\ref{BD1}) and (\ref{BD2}) with an additional
factor of the light-cone momentum fraction $x=k^+/P^+$ of the struck
constituent in the integrand.  The $B(q^2)$ form factor is obtained from
the non-forward spin-flip amplitude.  The value of $B(0)$ is obtained in
the $q^2 \to 0$ limit.
The angular
momentum projection of a state is given by
\begin{equation}
\VEV{J^i} = {1\over 2} \epsilon^{i j k} \int d^3x \VEV{T^{0 k}x^j - T^{0 j} x^k}
= A(0) \VEV{L^i} + \left[A(0) + B(0)\right] \bar u(P){1\over
2}\sigma^i u(P)
\ .
\label{Ji13a}
\end{equation}
This result is derived using a wave-packet description of the state.  The
$\VEV{L^i}$ term is the orbital angular momentum of the center of  mass motion
with respect to an arbitrary origin and can be dropped.  The coefficient
of the $\VEV{L^i}$ term must be 1; $A(0) = 1 $ also follows  when we evaluate
the four-momentum expectation value $\VEV{P^\mu}$.
Thus the total intrinsic angular momentum
$J^z$ of a nucleon can be identified with the values of the form factors
$A(q^2)$ and
$B(q^2)$ at
$q^2= 0:$
\begin{equation}
      \VEV{J^z} = \VEV{{1\over 2} \sigma^z} \left[A(0) + B(0)\right] \ .
\label{Ji13}
\end{equation}

One  can  define
individual  quark and gluon contributions to the total
angular momentum from the matrix elements of the energy
momentum tensor.\cite{Ji97}
However, this definition is only formal; $A_{q,g}(0)$ can be interpreted
as the light-cone momentum fraction carried by the quarks or gluons
$\VEV{x_{q,g}}.$ The contributions from $ B_{q,g}(0) $ to $J_z$ cancel in the
sum.  In fact, it will be shown below \cite{Brodsky:2000ii}  that
the contributions to $B(0)$ vanish when summed over the constituents of
each individual Fock state.

We will give an explicit realization of
these relations in the light-cone Fock
representation for general composite systems.   In the next section we
will illustrate the formulae by computing the
electron's electromagnetic and energy-momentum tensor form factors to
one-loop order in QED.  In fact, the structure of this calculation has
much more generality and can be used as a template for more general
composite systems.

\section{ The Light-Cone Fock State Decomposition and Spin Structure of
Leptons in QED}

Recently Dae Sung Hwang, Bo-Qiang Ma, Ivan Schmidt, and
I \cite{Brodsky:2000ii} have shown that the light-cone wavefunctions
generated by the radiative corrections to the electron in QED provides a
simple system for understanding the spin and angular momentum
decomposition of relativistic systems.  This
perturbative model also illustrates the interconnections between Fock
states of different number.  The model is patterned after the quantum
structure which occurs in the one-loop Schwinger
${\alpha / 2 \pi} $ correction to the electron magnetic
moment.\cite{Brodsky:1980zm} In effect, we can represent a spin-$\half$ ~
system as a composite of a spin-$\half$ ~ fermion and spin-one vector
boson with arbitrary masses.  A similar model has been used to
illustrate the matrix elements and evolution of light-cone helicity and
orbital angular momentum operators.\cite{Harindranath:1999ve} This
representation of a composite system is particularly useful because it
is based on two constituents but yet is totally relativistic.  We can
then explicitly compute the form factors
$F_1(q^2)$ and $F_2(q^2)$ of the electromagnetic current, and the
various contributions to the form factors
$A(q^2)$ and $B(q^2)$ of the energy-momentum tensor.  The
anomalous moment coupling $B(0)$ to a graviton is shown to vanish for
any composite system.  This remarkable
result, first derived by Okun and Kobzarev,
\cite{Okun,Ji:1996kb,Ji:1997ek,Ji:1997nm,Teryaev:1999su} is shown to
follow directly from the Lorentz
boost properties of the light-cone Fock
representation.\cite{Brodsky:2000ii}

The Schwinger one-loop radiative correction to the electron current in
quantum electrodynamics has played a historic role in the development of
quantum field theory.  In the language of light-cone quantization, the
electron anomalous magnetic moment
$a_e = {\alpha/ 2 \pi}$ is due to the one-fermion one-gauge boson
Fock state component of the physical electron.  An explicit calculation of
the anomalous moment in this framework was given by Brodsky and
Drell.\cite {BD80}  We shall show here that the light-cone wavefunctions
of the electron provides an ideal system to check explicitly the
intricacies of spin and angular momentum in quantum field theory.  In
particular, we shall evaluate the matrix elements of the QED energy
momentum tensor and show how the ``spin crisis" is resolved in QED for
an actual physical system.  The analysis is exact in perturbation theory.
The same method can be applied to the moments of structure functions and
the evaluation of other local matrix elements.
We will also show how the
perturbative light-cone wavefunctions of leptons and photons provide a
template for the wavefunctions of non-perturbative composite systems
resembling hadrons in QCD.

The light-cone Fock state wavefunctions of an electron can be
systematically evaluated in QED.  The QED Lagrangian density is
\begin{equation}
{\cal{L}}={i\over 2}\ [\ {\bar{\psi}}\gamma^\mu
(\overrightarrow{\partial}{}_\mu +ieA_\mu )\psi
-{\bar{\psi}}\gamma^\mu
(\overleftarrow{\partial}{}_\mu -ieA_\mu )\psi \ ]-m{\bar{\psi}}\psi
-{1\over 4}F^{\mu\nu}F_{\mu\nu} \ ,
\label{fmm1v}
\end{equation}
and the corresponding energy-momentum tensor is
\begin{eqnarray}
T^{\mu\nu}&=&{i\over 4}\ \Big(\
[\ {\bar{\psi}}\gamma^\mu (\overrightarrow{\partial}{}^\nu +ieA^\nu)\psi
-{\bar{\psi}}\gamma^\mu (\overleftarrow{\partial}{}^\nu  -ieA^\nu)\psi \ ]
\ +\ [\ \mu\longleftrightarrow\nu\ ]\ \Big)
\nonumber\\
&&+\ F^{\mu\rho}F_{\rho}^{\ \nu}
\ +\ {1\over 4}g^{\mu\nu}F^{\rho\lambda}F_{\rho\lambda} \ .
\label{lagsv}
\end{eqnarray}
Since $T^{\mu\nu}$ is the  Noether current of the general
coordinate transformation, it is conserved.  In later calculations
we will identify the two terms in Eq. (\ref{lagsv}) as the fermion and
boson contributions
$T^{\mu\nu}_{\rm f}$  and $T^{\mu\nu}_{\rm b}$, respectively.

The physical electron is the eigenstate of the QED
Hamiltonian.  As  discussed in the introduction, the expansion of it is
the QED eigenfunction on the complete set $\left|n\right>$ of $H_0$
eigenstates produces the Fock state expansion.  It is particularly
advantageous to carry out this procedure using light-cone quantization
since the vacuum is trivial, the Fock state representation is boost
invariant, and the light-cone fractions $x_i = k^+_i/P^+$ are positive:
$0 < x_i  \le 1$,
$\sum_i x_i = 1$.  We also employ light-cone gauge $A^+ = 0$ so that the
gauge boson polarizations are physical.
Thus each Fock-state wavefunction $\left< n |{\rm physical \
electron} \right> $ of the physical electron with total spin projection
$J^z = \pm {1\over 2}$ is represented by the function
$\psi^{J^z}_n(x_i,{\vec k}_{\perp i},\lambda_i)$, where
\begin{equation}
k_i=(k^+_i,k^-_i,{\vec k}_{ \perp i})= \left(x_i P^+, \frac{{\vec
k}_{\perp i}^2+m_i^2}{x_i P^+}, {\vec k}_{\perp i}\right)
\end{equation}
specifies the momentum of each constituent and $\lambda_i$ specifies
its light-cone helicity in the $z$ direction.  We  adopt a non-zero
boson mass $\lambda$ for the sake of generality.

The two-particle Fock state for an electron with $J^z = + {1\over 2}$ has
four possible spin combinations:
\begin{eqnarray}
&&\left|\Psi^{\uparrow}_{\rm two \ particle}(P^+=1, \vec P_\perp = \vec
0_\perp)\right> \qquad\qquad
\label{vsn1}\\
&=&
\int\frac{{\mathrm d}^2 {\vec k}_{\perp} {\mathrm d} x }
{16 \pi^3 \sqrt{x(1-x}}
\Bigg[ \
\psi^{\uparrow}_{+\frac{1}{2}\, +1}(x,{\vec k}_{\perp})\,
\left| +\frac{1}{2}\, +1\, ;\,\, x\, ,\,\, {\vec k}_{\perp}\right>
\nonumber \\
&&\quad+\psi^{\uparrow}_{+\frac{1}{2}\, -1}(x,{\vec k}_{\perp})\,
\left| +\frac{1}{2}\, -1\, ;\,\, x\, ,\,\, {\vec k}_{\perp}\right>
\nonumber\\
&&\quad +\psi^{\uparrow}_{-\frac{1}{2}\, +1} (x,{\vec k}_{\perp})\,
\left| -\frac{1}{2}\, +1\, ;\,\, x\, ,\,\, {\vec k}_{\perp}\right>
\nonumber \\
&&\quad +\psi^{\uparrow}_{-\frac{1}{2}\, -1} (x,{\vec k}_{\perp})\,
\left| -\frac{1}{2}\, -1\, ;\,\, x\, ,\,\, {\vec k}_{\perp}\right>\ \Bigg]
\ .
\nonumber
\end{eqnarray}
The wavefunctions can be evaluated explicitly in QED
perturbation theory using the rules given by Brodsky and Lepage \cite{BL80}
and Brodsky and Drell \cite{BD80}:
\begin{equation}
\left
\{ \begin{array}{l}
\psi^{\uparrow}_{+\frac{1}{2}\, +1} (x,{\vec k}_{\perp})=-{\sqrt{2}}
\frac{(-k^1+{\mathrm i} k^2)}{x(1-x)}\,
\varphi \ ,\\
\psi^{\uparrow}_{+\frac{1}{2}\, -1} (x,{\vec k}_{\perp})=-{\sqrt{2}}
\frac{(+k^1+{\mathrm i} k^2)}{1-x }\,
\varphi \ ,\\
\psi^{\uparrow}_{-\frac{1}{2}\, +1} (x,{\vec k}_{\perp})=-{\sqrt{2}}
(M-{m\over x})\,
\varphi \ ,\\
\psi^{\uparrow}_{-\frac{1}{2}\, -1} (x,{\vec k}_{\perp})=0\ ,
\end{array}
\right.
\label{vsn2}
\end{equation}
where
\begin{equation}
\varphi=\varphi (x,{\vec k}_{\perp})=\frac{ e/\sqrt{1-x}}{M^2-({\vec
k}_{\perp}^2+m^2)/x-({\vec k}_{\perp}^2+\lambda^2)/(1-x)}\ .
\label{wfdenom}
\end{equation}

Similarly,
\begin{eqnarray}
&&\left|\Psi^{\downarrow}_{\rm two \ particle}(P^+=1, \vec P_\perp =
\vec 0_\perp)\right>
\label{vsn1a}\\
&=&
\int\frac{{\mathrm d}^2 {\vec k}_{\perp} {\mathrm d} x }
{16 \pi^3 \sqrt{x(1-x}}
\Bigg[\
\psi^{\downarrow}_{+\frac{1}{2}\, +1}(x,{\vec k}_{\perp})\,
\left| +\frac{1}{2}\, +1\, ;\,\, x\, ,\,\, {\vec k}_{\perp}\right>
\nonumber \\
&&\quad +\psi^{\downarrow}_{+\frac{1}{2}\, -1}(x,{\vec k}_{\perp})\,
\left| +\frac{1}{2}\, -1\, ;\,\, x\, ,\,\, {\vec k}_{\perp}\right>
\nonumber\\
&&\quad +\psi^{\downarrow}_{-\frac{1}{2}\, +1}(x,{\vec k}_{\perp})\,
\left| -\frac{1}{2}\, +1\, ;\,\, x\, ,\,\, {\vec k}_{\perp}\right>
\nonumber \\
&&\quad+\psi^{\downarrow}_{-\frac{1}{2}\, -1}(x,{\vec k}_{\perp})\,
\left| -\frac{1}{2}\, -1\, ;\,\, x\, ,\,\, {\vec k}_{\perp}\right>\ \Bigg]
\ ,
\nonumber
\end{eqnarray}
where
\begin{equation}
\left
\{ \begin{array}{l}
\psi^{\downarrow}_{+\frac{1}{2}\, +1} (x,{\vec k}_{\perp})=0\ ,\\
\psi^{\downarrow}_{+\frac{1}{2}\, -1} (x,{\vec k}_{\perp})=-{\sqrt{2}}
(M-{m\over x})\,
\varphi \ ,\\
\psi^{\downarrow}_{-\frac{1}{2}\, +1} (x,{\vec k}_{\perp})=-{\sqrt{2}}
\frac{(-k^1+{\mathrm i} k^2)}{1-x }\,
\varphi \ ,\\
\psi^{\downarrow}_{-\frac{1}{2}\, -1} (x,{\vec k}_{\perp})=-{\sqrt{2}}
\frac{(+k^1+{\mathrm i} k^2)}{x(1-x)}\,
\varphi \ .
\end{array}
\right.
\label{vsn2a}
\end{equation}
The  coefficients of $\varphi$ in  Eqs.  (\ref{vsn2})
and (\ref{vsn2a}) are the matrix elements of
$\frac{\overline{u}(k^+,k^-,{\vec k}_{\perp})}{{\sqrt{k^+}}}
\gamma \cdot \epsilon^{*}
\frac{u (P^+,P^-,{\vec P}_{\perp})}{{\sqrt{P^+}}}$
which are
the numerators of the wavefunctions corresponding to
each constituent spin $s^z$ configuration.
The two boson polarization vectors in light-cone gauge are $\epsilon^\mu=
(\epsilon^+ = 0\ , \epsilon^- = {\vec \epsilon_\perp \cdot \vec k_\perp
\over 2 k^+}, \vec
\epsilon_\perp)$ where
$\vec{\epsilon}=\vec{\epsilon_\perp}_{\uparrow,\downarrow}=
\mp (1/\sqrt{2})(\widehat{x} \pm {\mathrm i} \widehat{y})$.
The polarizations also satisfy the Lorentz condition
$  k \cdot \epsilon =0$.

Note that each Fock state configuration satisfies
the spin sum rule: $J^z=S^z_{\rm f}+s^z_{\rm b} + l^z = +{1\over 2}$.
The sign of the helicity of the
electron is retained by the leading photon at $x_\gamma = 1- x \to 1$.
Note that in the non-relativistic limit, the transverse motion of the
constituents can be neglected, and we have only the
$\left|+\frac{1}{2}\right> \to
\left|-\frac{1}{2}\, +1\right>$ configuration which is the
non-relativistic quantum state for the spin-half system composed of
a fermion and a spin-1 boson constituents.  The fermion
constituent has spin projection in the opposite
direction to the spin $J^z$ of the whole system.
However, for ultra-relativistic binding in which
the transversal motions of the constituents are large compared to the
fermion masses,  the
$\left|+\frac{1}{2}\right> \to \left|+\frac{1}{2}\, +1\right>$
and
$\left|+\frac{1}{2}\right> \to \left|+\frac{1}{2}\, -1\right>$
configurations dominate
over the $\left|+\frac{1}{2}\right> \to \left|-\frac{1}{2}\, +1\right>$
configuration.  In this case
the fermion constituent has
spin projection parallel to $J^z$.

We can see
how the angular momentum sum rule is satisfied  for
the wavefunctions Eqs. (\ref{vsn1}) and (\ref{vsn1a}) of the QED model
system.  In Table~1 we list the fermion constituent's light-cone spin
projection
$s^z_{\rm f} = {1\over 2} \lambda_{\rm f}$, the boson constituent spin
projection
$s^z_{\rm b} = \lambda_{\rm b}$, and the relative orbital angular momentum
$l^z$ for each contributing configuration of the QED model system
wavefunction.
\begin{table}[ht]
\begin{center}
Table 1.  Spin Decomposition of the $J^z_e = + 1/2$ Electron  \\
\vspace{.3truecm}
\begin{footnotesize}
\begin{tabular}{|c|c|c|c|}
\hline\ru1
&  Fermion & Boson & Orbital Ang. \\[-2ex]
$\ \ \ \ $Configuration$\ \ \ \ $ & $\ \ \ $ Spin $s^z_{\rm f}$$\ \ \ $
&$\ \ \ $  Spin $s^z_{\rm b}$$\ \ \ $
&   Mom. $l^z$\\[1ex]
\hline\hline\ru1
{$\left|+\frac{1}{2}\right> \to  \left|+\frac{1}{2}\, +1\right> $}
& {$+{1\over 2}$}
& {$+1$}
& {$-1$}\\
\hline\ru1
{$\left|+\frac{1}{2}\right> \to  \left|-\frac{1}{2}\, +1\right> $}
& {$-{1\over 2}$}
& {$+1$}
& {$0$}\\
\hline\ru1
{$\left|+\frac{1}{2}\right> \to  \left|+\frac{1}{2}\, -1\right> $}
& {$+{1\over 2}$}
& {$-1$}
& {$+1$}\\
\hline
\end{tabular}
\end{footnotesize}
\end{center}
\end{table}
Table~1 
is derived by
calculating the matrix elements of the light-cone helicity
operator $\gamma^+\gamma^5$ \cite{Ma91b}
and the relative orbital angular momentum
operator $-{\mathrm i} (k^1\frac{\partial}{\partial k^2}
-k^2\frac{\partial}{\partial k^1})$ \cite{Harindranath:1999ve,MS98,Hag98}
in the light-cone representation.  Each configuration satisfies
the spin sum rule: $J^z=s^z_{\rm f}+s^z_{\rm b} + l^z$.

The electron in QED also has a ``bare" one-particle component:
\begin{equation}
|\Psi_{\rm one \ particle}^{\uparrow , \downarrow}\rangle =
\sqrt Z \,\, \delta(1-x) \,\,\delta(\vec k_\perp = \vec 0_\perp )\,\,
\delta_{s^z_{\rm f} \  \pm {1\over 2}} \ ,
\label{bare}
\end{equation}
where $Z$ is the wavefunction normalization of the one-particle
state.  If we regulate the theory in the ultraviolet and infrared, $Z$ is
finite.

We first will evaluate the Dirac and Pauli form factors $F_1(q^2)$ and
$F_2(q^2)$.  Using  Eqs. (\ref{BD1}) and (\ref{vsn1}) we
have to order $e^2$
\begin{eqnarray}
F_1(q^2)&=&
\left<\Psi^{\uparrow}(p^+=1, {\vec P_\perp}= \vec q_\perp))
|\Psi^{\uparrow}(p^+=1, {\vec P_\perp}= \vec 0_\perp)\right>
\nonumber\\
&=& Z + \int\frac{{\mathrm d}^2 {\vec k}_{\perp} {\mathrm d} x }{16 \pi^3}
\Big[\psi^{\uparrow\ *}_{+\frac{1}{2}\, +1}(x,{\vec k'}_{\perp})
\psi^{\uparrow}_{+\frac{1}{2}\, +1}(x,{\vec k}_{\perp})
 \\
&& \quad+\psi^{\uparrow\ *}_{+\frac{1}{2}\, -1}(x,{\vec k'}_{\perp})
\psi^{\uparrow}_{+\frac{1}{2}\, -1}(x,{\vec k}_{\perp})
+\psi^{\uparrow\ *}_{-\frac{1}{2}\, +1}(x,{\vec k'}_{\perp})
\psi^{\uparrow}_{-\frac{1}{2}\, +1}(x,{\vec k}_{\perp})
\Big]\ ,
\label{BDF1a}\nonumber
\end{eqnarray}
where
\begin{equation}
{\vec k'}_{\perp}={\vec k}_{\perp}+(1-x){\vec q}_{\perp}\ .
\label{BDF1b}
\end{equation}
Ultraviolet regularization is assumed.  For example, we can assume a
cutoff in the invariant mass of the constituents:
${\cal M}^2 = \sum_i {{\vec k}^2_{\perp i} +m^2_i\over x_i} < \Lambda^2.$

At zero momentum transfer
\begin{eqnarray}
F_1(0)&=& Z
+\int\frac{{\mathrm d}^2 {\vec k}_{\perp} {\mathrm d} x }{16 \pi^3}
\Big[ \psi^{\uparrow\ *}_{+\frac{1}{2}\, +1}(x,{\vec k}_{\perp})
\psi^{\uparrow}_{+\frac{1}{2}\, +1}(x,{\vec k}_{\perp})
\\ &&\quad
+\psi^{\uparrow\ *}_{+\frac{1}{2}\, -1}(x,{\vec k}_{\perp})
\psi^{\uparrow}_{+\frac{1}{2}\, -1}(x,{\vec k}_{\perp})
+\psi^{\uparrow\ *}_{-\frac{1}{2}\, +1}(x,{\vec k}_{\perp})
\psi^{\uparrow}_{-\frac{1}{2}\, +1}(x,{\vec k}_{\perp})
\Big]\ . \nonumber
\label{BDF1}
\end{eqnarray}
We can simulate a composite model of two particles by
choosing the coupling strength $e(\Lambda)$ such that $F_1(0) = 1$ is
satisfied.  The one-loop model can
be further generalized by applying spectral Pauli-Villars
integration over the constituent masses.  The resulting
form of light-cone wavefunctions provides a template
for parameterizing the structure of relativistic composite systems and
their matrix elements in hadronic physics.

The Pauli form factor is obtained from the spin-flip matrix element of
the $J^+$ current.   From Eqs. (\ref{BD2}), (\ref{vsn1}), and (\ref{vsn1a})
we have
\begin{eqnarray}
F_2(q^2)&=&
{-2M\over (q^1-{\mathrm i}q^2)}
\left <\Psi^{\uparrow}(P^+=1, {\vec P_\perp}= \vec q_\perp))
|\Psi^{\downarrow}(P^+=1, {\vec P_\perp}= \vec 0_\perp)\right>
\nonumber\\
&=&{-2M\over (q^1-{\mathrm i}q^2)}
\int\frac{{\mathrm d}^2 {\vec k}_{\perp} {\mathrm d} x }{16 \pi^3}
\Big[\psi^{\uparrow\ *}_{+\frac{1}{2}\, -1}(x,{\vec k'}_{\perp})
\psi^{\downarrow}_{+\frac{1}{2}\, -1}(x,{\vec k}_{\perp})
\nonumber \\
&& +\psi^{\uparrow\ *}_{-\frac{1}{2}\, +1}(x,{\vec k'}_{\perp})
\psi^{\downarrow}_{-\frac{1}{2}\, +1}(x,{\vec k}_{\perp})
\Big]
\nonumber\\
&=&4M\int\frac{{\mathrm d}^2 {\vec k}_{\perp} {\mathrm d} x }{16 \pi^3}
{(m-Mx)\over x}\varphi (x,{\vec k'}_{\perp}){}^*
\varphi (x,{\vec k}_{\perp})
\nonumber\\
&=&4Me^2\int\frac{{\mathrm d}^2 {\vec k}_{\perp} {\mathrm d} x }{16 \pi^3}
{(m-xM)\over x(1-x)}
\nonumber\\
&&\times \frac{1}
{\left[M^2-
 \frac{({\vec k}_{\perp}+(1-x){\vec q}_{\perp})^2+m^2}         {x}
-\frac{({\vec k}_{\perp}+(1-x){\vec q}_{\perp})^2+\lambda^2}{1-x}\right]
}
\nonumber\\[2ex]
&&\times \frac{1}
{\left[{M^2-
\frac{{\vec k}_{\perp}^2+m^2}{x}
-\frac{{\vec k}_{\perp}^2+\lambda^2}{1-x}}\right]
}\ .
\label{BDF2a}
\end{eqnarray}
Using the Feynman parameterization, we can also express Eq. (\ref{BDF2a})
in a form in which the $q^2=-{\vec q}_{\perp}^2$ dependence is more
explicit as
\begin{equation}
F_2(q^2)=
{Me^2\over 4\pi^2}\int_0^1d\alpha\,\int_0^1dx\,\,
{m-xM\over
\alpha (1-\alpha)\, {1-x\over x}\, {\vec q}_{\perp}^2-M^2
+{m^2\over x}+{\lambda^2\over 1-x}}\ .
\label{BDF2az}
\end{equation}

The anomalous moment is obtained in the limit of zero momentum transfer:
\begin{eqnarray}
F_2(0)&=&
4Me^2\int\frac{{\mathrm d}^2 {\vec k}_{\perp} {\mathrm d} x }{16 \pi^3}
{(m-xM) \over x(1-x)}\,\,
\frac{1}{
\left[M^2-
\frac{{\vec k}_{\perp}^2+m^2}{x}-
\frac{{\vec k}_{\perp}^2+\lambda^2}{1-x}\right]^2}
\nonumber\\
&=&{Me^2\over 4\pi^2}\int_0^1dx\,\,
{m-xM\over
-M^2
+{m^2\over x}+{\lambda^2\over 1-x}}
\ ,
\label{BDF2z}
\end{eqnarray}
which is the result of Brodsky and Drell.\cite{BD80}
For zero photon mass and $M=m$, it gives the correct order $\alpha$
Schwinger value $a_e=F_2(0)={\alpha / 2\pi}$ for the electron anomalous
magnetic moment for QED.

As seen from Eqs. (\ref{eBD1}) and (\ref{eBD2}), the matrix elements of the
double plus
components of the energy-momentum tensor are sufficient to derive the
fermion and boson constituents' form factors
$A_{\rm f,g}(q^2)$ and $B_{\rm f,g}(q^2)$ of graviton coupling to matter.
In particular, we
shall verify $A(0)=A_{\rm f}(0)+A_{\rm b}(0) =1$ and $B(0) = 0 .$

The individual contributions of the fermion and boson fields
to the energy-momentum form factors in QED are given by
\begin{eqnarray}
A_{\rm f}(q^2)&=&
\left<\Psi^{\uparrow}(P^+=1,{\vec P_\perp}={\vec q_\perp})
\right|\frac{T^{++}_{\rm f}(0)}{2(P^+)^2}
\left|\Psi^{\uparrow}(P^+=1,{\vec
P_\perp}={\vec 0_\perp})\right>
\nonumber\\
&=&\int\frac{{\mathrm d}^2 {\vec k}_{\perp} {\mathrm d} x }{16 \pi^3}
\ x\
\Big[\psi^{\uparrow\ *}_{+\frac{1}{2}\, +1}(x,{\vec k'}_{\perp})
\psi^{\uparrow}_{+\frac{1}{2}\, +1}(x,{\vec k}_{\perp})
\\
&&\quad
+\psi^{\uparrow\ *}_{+\frac{1}{2}\, -1}(x,{\vec k'}_{\perp})
\psi^{\uparrow}_{+\frac{1}{2}\, -1}(x,{\vec k}_{\perp})
+\psi^{\uparrow\ *}_{-\frac{1}{2}\, +1}(x,{\vec k'}_{\perp})
\psi^{\uparrow}_{-\frac{1}{2}\, +1}(x,{\vec k}_{\perp})
\Big]\ , \nonumber
\label{BDF1ae}
\end{eqnarray}
where
${\vec k'}_{\perp}$ is given in Eq. (\ref{BDF1b}),
and
\begin{eqnarray}
A_{\rm b}(q^2)&=&
\left<\Psi^{\uparrow}(P^+=1,{\vec P_\perp}={\vec q_\perp})
\right|\frac{T^{++}_{\rm b}(0)}{2(P^+)^2}
\left|\Psi^{\uparrow}(P^+=1,{\vec P_\perp}={\vec
0_\perp})\right>
\nonumber\\
&=&\int\frac{{\mathrm d}^2 {\vec k}_{\perp} {\mathrm d} x }{16 \pi^3}
\ (1-x)\
\Big[\psi^{\uparrow\ *}_{+\frac{1}{2}\, +1}(x,{\vec k''}_{\perp})
\psi^{\uparrow}_{+\frac{1}{2}\, +1}(x,{\vec k}_{\perp})
\\
&& \quad +\psi^{\uparrow\ *}_{+\frac{1}{2}\, -1}(x,{\vec k''}_{\perp})
\psi^{\uparrow}_{+\frac{1}{2}\, -1}(x,{\vec k}_{\perp})
+\psi^{\uparrow\ *}_{-\frac{1}{2}\, +1}(x,{\vec k''}_{\perp})
\psi^{\uparrow}_{-\frac{1}{2}\, +1}(x,{\vec k}_{\perp})
\Big]\ , \nonumber
\label{BDF1ag}
\end{eqnarray}
where
\begin{equation}
{\vec k''}_{\perp}={\vec k}_{\perp}-x{\vec q}_{\perp}\ .
\label{BDF1bg}
\end{equation}
Note that
\begin{equation}
A_{\rm f}(0)+A_{\rm b}(0)=F_1(0) = 1\ ,
\label{BDF1ah}
\end{equation}
which corresponds to the momentum sum rule.

The fermion and boson contributions to the spin-flip matter form factor
are
\begin{eqnarray}
B_{\rm f}(q^2)&=&
{-2M\over (q^1-{\mathrm i}q^2)}
\left<\Psi^{\uparrow}(P^+=1,{\vec P_\perp}={\vec q_\perp})
\right|\frac{T^{++}_{\rm f}(0)}{2(P^+)^2}
\left|\Psi^{\downarrow}(P^+=1,{\vec P_\perp}={\vec
0_\perp})\right>
\nonumber\\
&=&{-2M\over (q^1-{\mathrm i}q^2)}
\int\frac{{\mathrm d}^2 {\vec k}_{\perp} {\mathrm d} x }{16 \pi^3}\ x\
\Big[\psi^{\uparrow\ *}_{+\frac{1}{2}\, -1}(x,{\vec
k'}_{\perp})
\psi^{\downarrow}_{+\frac{1}{2}\, -1}(x,{\vec k}_{\perp})
\nonumber \\
&& \quad +\psi^{\uparrow\ *}_{-\frac{1}{2}\, +1}(x,{\vec k'}_{\perp})
\psi^{\downarrow}_{-\frac{1}{2}\, +1}(x,{\vec k}_{\perp})
\Big]
\nonumber\\
&=&4M\int\frac{{\mathrm d}^2 {\vec k}_{\perp} {\mathrm d} x }{16 \pi^3}
(m-Mx)\varphi (x,{\vec k'}_{\perp}){}^*
\varphi (x,{\vec k}_{\perp})
\nonumber\\
&=&4Me^2\int\frac{{\mathrm d}^2 {\vec k}_{\perp} {\mathrm d} x }{16 \pi^3}
{(m-xM)\over (1-x)}
\nonumber\\
&&\times {1\over \left[
{M^2-\frac{({\vec k}_{\perp}+(1-x){\vec q}_{\perp})^2+m^2}{x}
-\frac{({\vec k}_{\perp}+(1-x){\vec q}_{\perp})^2+\lambda^2}{1-x}}\right]}
\nonumber\\[1ex]
&&\times {1\over
\left[M^2-\frac{{\vec k}_{\perp}^2+m^2}{x}-
\frac{{\vec k}_{\perp}^2+\lambda^2}{1-x}\right]}
\nonumber\\[1ex]
&=&{Me^2\over 4\pi^2}\int_0^1d\alpha\,\int_0^1dx\,\,
{x(m-xM)\over
\alpha (1-\alpha)\, {1-x\over x}\, {\vec q}_{\perp}^2-M^2
+{m^2\over x}+{\lambda^2\over 1-x}}
\ ,
\label{BDF2abe}
\end{eqnarray}
and
\begin{eqnarray}
B_{\rm b}(q^2)&=&
{-2M\over (q^1-{\mathrm i}q^2)}
\left<\Psi^{\uparrow}(P^+=1,{\vec P_\perp}={\vec q_\perp})
\right|\frac{T^{++}_{\rm b}(0)}{2(P^+)^2}
\left|\Psi^{\downarrow}(P^+=1,{\vec P_\perp}={\vec
0_\perp})\right>
\nonumber\\
&=&{-2M\over (q^1-{\mathrm i}q^2)}
\int\frac{{\mathrm d}^2 {\vec k}_{\perp} {\mathrm d} x }{16 \pi^3}\ (1-x)\
\nonumber\\
&&\qquad\qquad\times\Big[\psi^{\uparrow\ *}_{+\frac{1}{2}\, -1}(x,{\vec
k'}_{\perp})
\psi^{\downarrow}_{+\frac{1}{2}\, -1}(x,{\vec k}_{\perp})
+\psi^{\uparrow\ *}_{-\frac{1}{2}\, +1}(x,{\vec k'}_{\perp})
\psi^{\downarrow}_{-\frac{1}{2}\, +1}(x,{\vec k}_{\perp})
\Big]
\nonumber\\
&=&-4M\int\frac{{\mathrm d}^2 {\vec k}_{\perp} {\mathrm d} x }{16 \pi^3}
(m-Mx)\varphi (x,{\vec k'}_{\perp}){}^*
\varphi (x,{\vec k}_{\perp})
\nonumber\\
&=&-4Me^2\int\frac{{\mathrm d}^2 {\vec k}_{\perp} {\mathrm d} x }{16 \pi^3}
{(m-xM)\over (1-x)}
\nonumber\\
&&\times {1\over [{M^2-(({\vec k}_{\perp}-x{\vec q}_{\perp})^2+m^2)/x
-(({\vec k}_{\perp}-x{\vec q}_{\perp})^2+\lambda^2)/(1-x)}]}
\nonumber\\
&&\times {1\over
[{M^2-({\vec k}_{\perp}^2+m^2)/x-({\vec k}_{\perp}^2+\lambda^2)/(1-x)}]}
\nonumber\\
&=&-{Me^2\over 4\pi^2}\int_0^1d\alpha\,\int_0^1dx\,\,
{x(m-xM)\over
\alpha (1-\alpha)\, {x\over 1-x}\, {\vec q}_{\perp}^2-M^2
+{m^2\over x}+{\lambda^2\over 1-x}}
\ .
\label{BDF2abg}
\end{eqnarray}
The total contribution for general momentum transfer is
\begin{eqnarray}
&&B(q^2) = B_{\rm f}(q^2)+B_{\rm b}(q^2)
\nonumber\\
&=&
4Me^2\int\frac{{\mathrm d}^2 {\vec k}_{\perp} {\mathrm d} x }{16 \pi^3}
{(m-xM)\over (1-x)}
\nonumber\\
&&\times
{\{}
{1\over [{M^2-(({\vec k}_{\perp}+(1-x){\vec q}_{\perp})^2+m^2)/x
-(({\vec k}_{\perp}+(1-x){\vec q}_{\perp})^2+\lambda^2)/(1-x)}]}
\nonumber\\
&&\qquad -
{1\over [{M^2-(({\vec k}_{\perp}-x{\vec q}_{\perp})^2+m^2)/x
-(({\vec k}_{\perp}-x{\vec q}_{\perp})^2+\lambda^2)/(1-x)}]} {\}}
\nonumber\\
&&\times
{1\over
[{M^2-({\vec k}_{\perp}^2+m^2)/x-({\vec k}_{\perp}^2+\lambda^2)/(1-x)}]}
\nonumber\\
&=&{Me^2\over 4\pi^2}\int_0^1d\alpha\,\int_0^1dx\,\,
{x(m-xM)}
\label{BDF2abgs}\\
&&\times\Bigl( {1\over
\alpha (1-\alpha)\, {1-x\over x}\, {\vec q}_{\perp}^2-M^2
+{m^2\over x}+{\lambda^2\over 1-x}}
-{1\over
\alpha (1-\alpha)\, {x\over 1-x}\, {\vec q}_{\perp}^2-M^2
+{m^2\over x}+{\lambda^2\over 1-x}}\Bigr)
\ .
\nonumber
\end{eqnarray}
This is the analog of the Pauli form factor for a physical electron
scattering in a gravitational field and in general is not zero.  However at
zero momentum transfer
\begin{equation}
B(0) = B_{\rm f}(0)+B_{\rm b}(0)=0.
\label{BDF2abgt}
\end{equation}
This result
agree with the conclusions of Okun and Kobzarev \cite{Okun},  Ji
\cite{Ji97} and Teryaev.\cite{Teryaev:1999su}

\begin{figure}[htbp]
\leavevmode
\begin{center}
{\vskip-1in\epsfxsize=5in\epsfbox{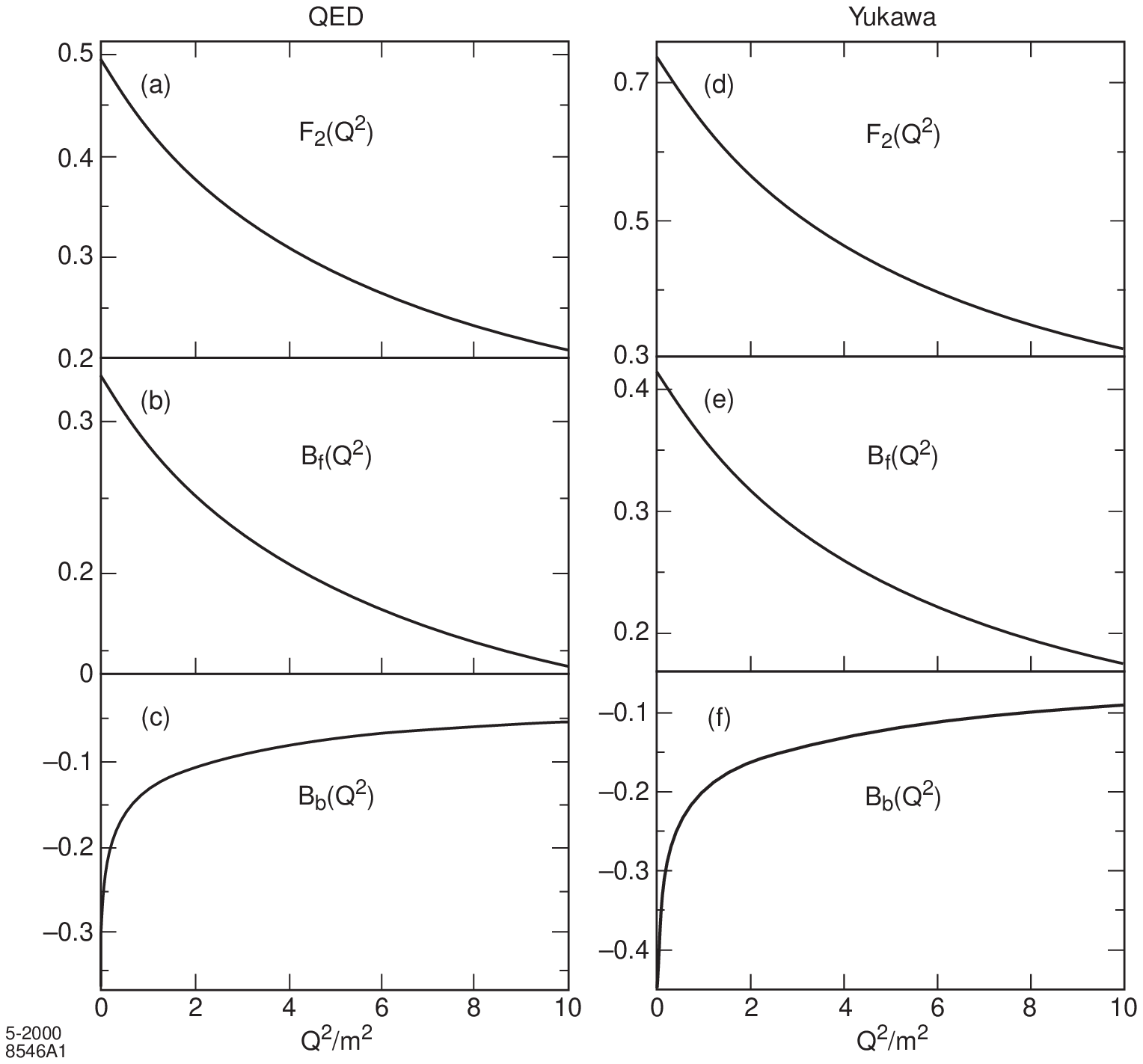}}
\end{center}
\vskip-1.25in
\caption[*]{Helicity-flip electromagnetic and gravitational form factors
for space-like $q^2=-Q^2<0$
from the quantum fluctuations of a fermion at one-loop order in units of
$\alpha/\pi$ for QED and $g^2/4\pi^2$ for the Yukawa theory.   The fermion
constituent mass is taken as $m_f=M$.  The boson constituent is massless.
(a)  Helicity-flip Pauli form factor $F_2(q^2)$ in QED.  Notice that
$F_2(0)=1/2$.
(b)  Helicity-flip form factor $B_b(q^2)$ of the graviton coupling
to  the boson (photon) constituent of the electron at one-loop order in
QED.  Notice that $B_b(0) = -1/3$.
(c)   Helicity-flip fermion form factor $B_f(q^2)$ of the graviton
coupling to the fermion constituent  at one-loop order in QED.  Notice
that $B_f(0)=1/3$, and thus $B_f(0) + B_b(0)=0.$
(d)  Helicity-flip Pauli form factor $F_2(q^2)$ in the Yukawa theory.
Notice that in this case $F_2(0)=3/4$.
(e)  Helicity-flip  form factor $B_b(q^2)$ of the graviton coupling
to  the boson at one-loop order in the Yukawa theory.  Notice that
$B_b(0)= - 5/12$.
(f)  Helicity-flip fermion form factor $B_f(q^2)$ of the graviton
coupling to the fermion constituent  at one-loop order in the Yukawa
theory.  Notice that $B_f(0)=5/12$, and thus $B_f(0) + B_b(0)=0.$}
\label{FIG1}
\end{figure}

The helicity-flip electromagnetic and gravitational form factors for the
fluctuations of the electron at one-loop are illustrated in Fig. \ref{FIG1}. The
cancelation of the sum of graviton couplings $B(q^2)$ to the constituents
at $q^2 = 0$ is evident.

\vspace{0.5cm}

\section {The Anomalous Gravitomagnetic Moment for Composite
Systems}

A remarkable property of gravitational interactions is that
the anomalous gravitomagnetic
moment $B(0)=0$ vanishes identically for each contributing Fock state of a
composite system.\cite{Brodsky:2000ii} In order to calculate $B(0)$ by
using Eq. (\ref{eBD2}), we need to consider a non-forward amplitude.
The internal momentum variables for the final state wavefunction are
given by Eqs. (\ref{kprime1}) and (\ref{kprime2}).
The subscripts of $x_i$ and $\vec k_{\perp i}$
label constituent particles, the superscripts of $q^1_\perp$, $k^1_\perp$,
and $k^2_\perp$ label the Lorentz indices, and the subscript $a$ in
$\psi_a$ indicates the contributing Fock state.  The essential ingredient
is the Lorentz property of the light-cone wavefunctions.

It is important to identify the $n-1$ independent relative momenta of
the $n$-particle Fock state.
\begin{eqnarray}
&&-{B(0)\over 2M}
=
\lim_{q_{\perp}^1\to 0}{\partial\over \partial q_{\perp}^1}
\VEV{P+q,\uparrow\left|\frac{T^{++}(0)}{2(P^+)^2}\right|P,\downarrow}
\label{gravito1}\\
&=&\lim_{q_{\perp}^1\to 0}{\partial\over \partial q_{\perp}^1}
\left<\Psi^{\uparrow}(P^+=1, {\vec P_\perp}= \vec
q_\perp))\left|\frac{T^{++}(0)}{2(P^+)^2}
\right|\Psi^{\downarrow}(P^+=1, {\vec
P_\perp}= \vec 0_\perp)\right>
\nonumber\\
&=&\lim_{q_{\perp}^1\to 0}{\partial\over \partial q_{\perp}^1}
\sum_{a}\int\prod^{n-1}_{k=1}
\frac{{\mathrm d}^2 {\vec k}_{\perp k} {\mathrm d} x_k }{16 \pi^3}
\nonumber \\
& \times &\psi^{\uparrow *}_a\Big( x_1,x_2,\cdots
,x_{n-1},(1-x_1-x_2-\cdots -x_{n-1}),
\nonumber\\
&&\qquad\qquad
{\vec k}_{\perp 1}',{\vec k}_{\perp 2}',\cdots ,{\vec k}_{\perp n-1}',
(-{\vec k}_{\perp 1}'-{\vec k}_{\perp 2}'-\cdots -{\vec k}_{\perp n-1}')\Big)
\nonumber\\
&\times &\Big[ \sum_{i=1}^{n-1}x_i+(1-x_1-x_2-\cdots -x_{n-1})\Big]
\nonumber\\
&\times &\psi^{\downarrow}_a
\Big( x_1,x_2,\cdots ,x_{n-1},(1-x_1-x_2-\cdots -x_{n-1}),
\nonumber\\
&&\qquad\qquad
{\vec k}_{\perp 1},{\vec k}_{\perp 2},\cdots ,{\vec k}_{\perp n-1},
(-{\vec k}_{\perp 1}-{\vec k}_{\perp 2}-\cdots
-{\vec k}_{\perp n-1})\Big) \ .
\nonumber
\end{eqnarray}

Using integration by parts,
\begin{eqnarray}
&&-{B_a(0)\over 2M}
=
\label{gravito2}\\
&=&\int\prod^{n-1}_{k=1}
\frac{{\mathrm d}^2 {\vec k}_{\perp k} {\mathrm d} x_k }{16 \pi^3}
\psi^{\uparrow *}_a\Big( x_1,x_2,\cdots ,x_{n-1},(1-x_1-x_2-\cdots -x_{n-1}),
\nonumber\\
&&\qquad\qquad
{\vec k}_{\perp 1},{\vec k}_{\perp 2},\cdots ,{\vec k}_{\perp n-1},
(-{\vec k}_{\perp 1}-{\vec k}_{\perp 2}-\cdots -{\vec k}_{\perp n-1})\Big)
\nonumber\\
&\times &
\Bigg[
\sum_{i=1}^{n-1}x_i\Big( (-1+x_i){\partial\over\partial k_{\perp i}^1}+
\sum_{j \ne i}^{n-1}x_j{\partial\over\partial k_{\perp j}^1}\Big)
\nonumber \\ & & \quad
+(1-x_1-x_2-\cdots -x_{n-1})
\sum_{j=1}^{n-1}x_j{\partial\over\partial k_{\perp j}^1}\Bigg]
\nonumber\\
&\times &\psi^{\downarrow}_a
\Big( x_1,x_2,\cdots ,x_{n-1},(1-x_1-x_2-\cdots -x_{n-1}),
\nonumber\\
&&\qquad\qquad
{\vec k}_{\perp 1},{\vec k}_{\perp 2},\cdots ,{\vec k}_{\perp n-1},
(-{\vec k}_{\perp 1}-{\vec k}_{\perp 2}-\cdots -{\vec k}_{\perp n-1})\Big)
\nonumber\\
&=&\int\prod^{n-1}_{k=1}
\frac{{\mathrm d}^2 {\vec k}_{\perp k} {\mathrm d} x_k }{16 \pi^3}
\psi^{\uparrow *}_a\Big( x_1,x_2,\cdots ,x_{n-1},(1-x_1-x_2-\cdots -x_{n-1}),
\nonumber\\
&&\qquad\qquad
{\vec k}_{\perp 1},{\vec k}_{\perp 2},\cdots ,{\vec k}_{\perp n-1},
(-{\vec k}_{\perp 1}-{\vec k}_{\perp 2}-\cdots -{\vec k}_{\perp n-1})\Big)
\nonumber\\
&\times &
\Big[
\sum_{i=1}^{n-1}
\Big( -1+\sum_{j=1}^{n-1}x_j+(1-x_1-x_2-\cdots -x_{n-1})\Big)
x_i{\partial\over\partial k_{\perp i}^1}\Big]
\nonumber\\
&\times &\psi^{\downarrow}_a
\Big( x_1,x_2,\cdots ,x_{n-1},(1-x_1-x_2-\cdots -x_{n-1}),
\nonumber\\
&&\qquad\qquad
{\vec k}_{\perp 1},{\vec k}_{\perp 2},\cdots ,{\vec k}_{\perp n-1},
(-{\vec k}_{\perp 1}-{\vec k}_{\perp 2}-\cdots -{\vec k}_{\perp n-1})\Big)
\nonumber\\
&=&0\ .
\nonumber
\end{eqnarray}
Thus the contribution $B_a(0)$ to
the total anomalous gravitomagnetic moment $B(0)$  vanishes separately
from each contributing Fock state $a$.

\section {The Perturbative Model as a Template for a Composite System}

The spin
structure of perturbative theory provides a
template for the numerator structure of the light-cone wavefunctions
even for composite systems since the equations which couple
different Fock components mimic the perturbative form.  For example,
the structure of the electron's Fock state in perturbative QED shows that
it is natural to have a negative contribution from relative orbital
angular momentum which balances the $S_z$ of its photon constituents.
We can thus expect a large orbital contribution to the proton's
$J_z$ since gluons carry roughly half of the proton's
momentum, thus providing insight into the ``spin crisis" in QCD.

We can generalize the perturbative model by using the structure of the
one-loop QED (and Yukawa) calculations with general values for general
values of the external mass $M$, internal fermion mass
$m$, and boson mass
$\lambda$, to represent a spin-$1\over 2$ system composed of a fermion
and a spin-1 or spin-0 boson.  Such a model describes an effectively
composite system with no bare one-particle Fock state.  We can also
generalize the functional form of the momentum space wavefunction
$\varphi(x,\vec k_\perp)$ by introducing a spectrum of vector
bosons satisfying the generalized Pauli-Villars spectral conditions
\begin{equation}
\int d \lambda^2  \lambda^{2N} \rho(\lambda^2) = 0, \ \ \ N= 0, 1, \cdots
\ .
\label{BDF2y}
\end{equation}
For example, we can simulate a proton
as a bound state of a quark and diquark \cite{Close}, using
spin-0, spin-1 diquarks, or a linear
superposition of the two states.
The model can be made to match the power-law fall-off of the hadron form
factors predicted in perturbative QCD by the choice of
sum rule conditions on the Pauli-Villars
spectra.\cite{Bro92,Brodsky:1998hs}   The light-cone framework of the
model resembles that of the covariant parton model of Landshoff,
Polkinghorne and Short
\cite{Landshoff:1971ff,Brodsky:1973hm}, in which the power behavior of
the spectral integral at high masses corresponds to the Regge behavior of
the deep inelastic structure functions.
Although the model is based  on just two Fock constituents,
it is relativistic and satisfies self-consistency
conditions such as in the point-like limit where
$R^2 M^2 \to 0$, the anomalous moment vanishes.\cite{Bro94}
The light-cone formalism
also properly incorporates Wigner boosts.  Thus this model of composite
systems can serve as a useful theoretical laboratory to interrelate
hadronic properties and check the consistency of formulae proposed for the
study of hadron substructure.

In the case of Yukawa theory at one loop, the non-relativistic fermion's
spin projection is aligned with the total $J^z$, and it is anti-aligned
in the ultra-relativistic limit.
The distinct features of spin structure
in the non-relativistic and ultra-relativistic
limits reveals the importance
of relativistic effects and
supports the viewpoint \cite{Ma91b,Ma96,MSS97}
that the proton ``spin
puzzle" can be understood as due to the relativistic
motion of quarks inside the nucleon.  In particular, the spin projection
of the relativistic constituent quark tends to be anti-aligned with the
proton spin in a quark-diquark bound state if the diquark has spin 0.  The
state with orbital angular momentum $l^z= \pm 1 $ in fact dominates over
the states with $l^z = 0.$   Thus the empirical fact that $\Delta q$ is
small in the proton has a natural description in the light-cone Fock
representation of hadrons.

\section{Light-cone Representation of Deeply Virtual 
\hfill\break Compton Scattering}

\begin{figure}[htb]
\begin{center}
\leavevmode
{\epsfxsize=3in\epsfbox{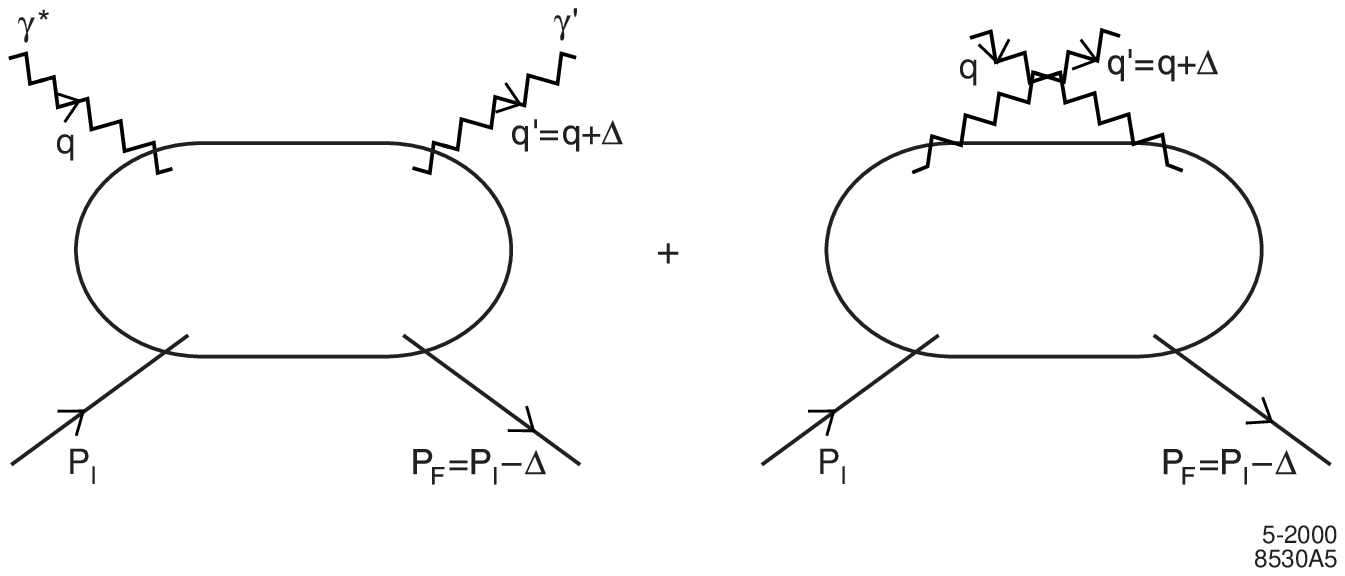}}
\end{center}
\caption[*]{The virtual Compton amplitude $\gamma^*(q) p_I \to
\gamma(q^\prime) p_F$.}
\label{fig:1}
\end{figure}

The virtual Compton scattering process ${d\sigma\over dt}(\gamma^*
p \to \gamma p)$ for large initial photon virtuality
$q^2=-Q^2$ (see Fig.~\ref{fig:1}) has extraordinary sensitivity to
fundamental features of the proton's structure.  Even though the final
state photon is on-shell, the deeply virtual process probes the
elementary quark structure of the proton near the light cone as an
effective local current.  In contrast to deep inelastic scattering, which
measures only the absorptive part of the forward virtual Compton
amplitude $Im {\cal T}_{\gamma^* p \to
\gamma^* p}$, deeply virtual Compton scattering allows the measurement of
the phase and spin structure of proton matrix elements for general
momentum transfer squared $t$.  In addition, the interference of the
virtual Compton amplitude and Bethe-Heitler
wide angle scattering Bremsstrahlung amplitude where the photon is
emitted from the lepton line leads to an electron-positron asymmetry in
the ${e^\pm  p \to e^\pm \gamma p}$ cross section which is proportional
to the real part of the Compton
amplitude.\cite{BCG7273}  The deeply virtual Compton amplitude
$\gamma^* p \to \gamma p$ is related by crossing to another important
process
$\gamma^* \gamma
\to $ hadron pairs at fixed invariant mass which can be measured
in electron-photon collisions.\cite{Diehl:2000uv}

\vspace{.5cm}

To leading order in $1/Q$, the deeply virtual Compton scattering
amplitude factorizes as the convolution in $x$ of the
amplitude $t^{\mu \nu}$ for hard Compton scattering on a quark line with
the generalized Compton form factors $H(x,t,\zeta),$ $ E(x,t,\zeta)$,
$\tilde H(x,t,\zeta),$ and $\tilde E(x,t,\zeta)$  of the target
proton.\cite{Ji9697,Ji:1998pc,Radyushkin:1996nd,Ji:1998xh}
\cite{Guichon:1998xv,Vanderhaeghen:1998uc,Radyushkin:1999es,%
Collins:1999be,Diehl98,Diehl:1999tr,Blumlein:2000cx}
Here
$x$ is the light-cone momentum fraction of the struck quark, and
$\zeta= Q^2/2 P\cdot q$ plays the role of the Bjorken variable.  The
square of the four-momentum transfer from the proton is given by
\begin{equation}
t=\Delta^2\ = \ 2P\cdot \Delta\  =\
-{(\zeta^2M^2+{\vec \Delta_\perp}^2)\over (1-\zeta)}\ ,
\label{na1anew}
\end{equation}
where $\Delta$ is the difference of initial and final momenta of the
proton ($P=P'+\Delta$).
The form factor $H(x,t,\zeta)$ describes the proton response
when the helicity of the proton is unchanged, and
$E(x,t,\zeta)$ is for the case when the proton helicity is flipped.  Two
additional functions $\tilde H(x,t,\zeta),$ and $\tilde E(x,t,\zeta)$
appear, corresponding to the dependence of the Compton amplitude on
quark helicity.

\begin{figure}[htbp]
\begin{center}
\leavevmode
{\epsfxsize=4in\epsfbox{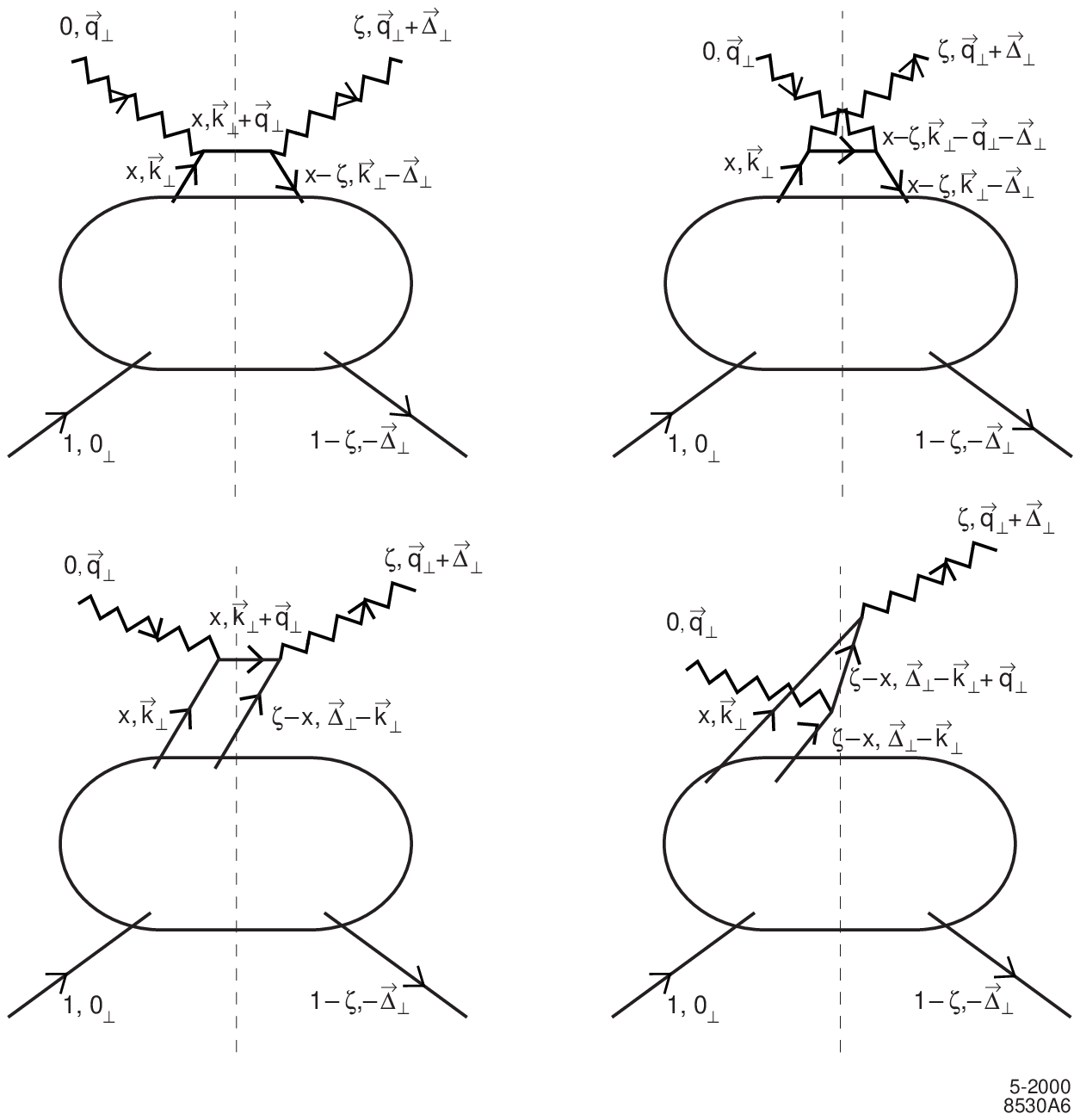}}
\end{center}
\caption[*]{Light-cone time-ordered contributions to deeply virtual
Compton scattering.  Only the contributions of leading twist in $1/q^2$
are illustrated.  These contributions illustrate the factorization
property of the leading twist amplitude.}
\label{fig:3}
\end{figure}

The kinematics of virtual Compton scattering
$\gamma^*(q) p(P) \to \gamma(q') p(P')$ are illustrated in
Fig.~\ref{fig:3}.  We specify the frame by choosing a convenient
parameterization of the light-cone coordinates for the initial and final
proton:
\begin{eqnarray}
P_I&=&(P^+\ ,\ {\vec P_\perp}\ ,\ P^-)\ =\ \left(\ P^+\ ,\ {\vec 0_\perp}\ ,\
{M^2\over P^+}\ \right)\ ,
\label{a1}\\
P_F&=&(P'^+\ ,\ {\vec P'_\perp}\ ,\ P'^-)\ =\
\left( (1-\zeta)P^+\ ,\ -{\vec \Delta_\perp}\ ,\ {(M^2+{\vec
\Delta_\perp}^2)\over (1-\zeta)P^+}\right)\ .
\nonumber\\
\end{eqnarray}
(Our metric is specified by $V^\pm = V^0 \pm V^z$ and $V^2 = V^+ V^- -
V_\perp^2$.) The four-momentum transfer from the
target is
\begin{eqnarray}
\Delta&=&P_I-P_F\ =\ (\Delta^+\ ,\ {\vec \Delta_\perp}\ ,\ \Delta^-)\ =\
\left( \zeta P^+\ ,\ {\vec \Delta_\perp}\ ,\
{(t+{\vec \Delta_\perp}^2)\over \zeta P^+}\right)\ ,
\nonumber
\end{eqnarray}
where $\Delta^2 = t$.
In addition, overall energy-momentum conservation requires $\Delta^- =
P^I_I - P^-_F.$

\vspace{.5cm}

As in the case of space-like form
factors, it is convenient to choose a frame where the incident space-like
photon carries $q^+ = 0$ and $q^2= - Q^2 = - {\vec q_\perp}^2$:
\begin{eqnarray}
q&=&(q^+\ ,\ {\vec q_\perp}\ ,\ q^-)\ =\ \left( 0\ ,\ {\vec q_\perp}\ ,\
{({\vec q_\perp}+{\vec \Delta_\perp})^2\over \zeta P^+}
+{(\zeta M^2+{\vec \Delta_\perp}^2)\over (1-\zeta)P^+}\right)\ ,
\\
q'&=&(q'^+\ ,\ {\vec q_\perp}^{\,\,\prime}\ ,\ q'^-)\ =\
\left( \zeta P^+\ ,\ ({\vec q_\perp}+{\vec \Delta_\perp})\ ,\
{({\vec q_\perp}+{\vec \Delta_\perp})^2\over \zeta P^+}\right)\
= q+\Delta .
\label{a2}\nonumber \\
\end{eqnarray}
Thus no light-cone time-ordered amplitudes involving the splitting of the
incident photon can  occur.  The connection between ${\vec
\Delta_\perp}^2$,
$\zeta$, and
$t$ is given by Eq. (\ref{na1anew}).
The variable $\zeta$ is fixed from (\ref{a1}) and (\ref{a2})
\begin{equation}
2P_I\cdot q={({\vec q_\perp}+{\vec \Delta_\perp})^2\over \zeta }
+{(\zeta M^2+{\vec \Delta_\perp}^2)\over (1-\zeta)}\ .
\label{nn2}
\end{equation}
We will be interested in deeply virtual Compton scattering where $q^2$ is
large compared to the masses and $t$.  Then, to leading order in $1/Q^2$,
we can take
\begin{equation}
{-q^2\over 2P_I\cdot q}=\zeta\ .
\label{nn3}
\end{equation}
Thus $\zeta$ plays the role of the Bjorken variable in deeply virtual
Compton scattering.
For a fixed value of $-t$, the allowed range of $\zeta$ is given by
\begin{equation}
0\ \le\ \zeta\ \le\
{(-t)\over 2M^2}\ \ \left( {\sqrt{1+{4M^2\over (-t)}}}\ -\ 1\ \right)\ .
\label{nn4}
\end{equation}
The choice of parameterization of the light-cone frame is of course
arbitrary.  For example, one can also
conveniently utilize a ``symmetric" frame for the ingoing and outgoing
proton which has manifest $\Delta \to - \Delta$ symmetry.

Recently, Markus Diehl, Dae Sung Hwang and I \cite{DVCSHwang} have shown
how the deeply virtual Compton amplitude can be evaluated explicitly in
the Fock state representation using the matrix elements of the currents
and the boost properties of the light-cone wavefunctions. For the $n \to
n$ diagonal term ($\Delta n = 0$), the arguments of the final-state hadron
wavefunction are
$x_1-\zeta \over 1-\zeta$,
${\vec{k}}_{\perp 1} - {1-x_1\over 1-\zeta} {\vec{\Delta}}_\perp$ for
the struck quark
and $x_i\over 1-\zeta$,
${\vec{k}}_{\perp i} + {x_i\over 1-\zeta} {\vec{\Delta}}_\perp$
for the $n-1$ spectators.
We thus obtain formulae for the diagonal (parton-number-conserving)
contribution to the generalized form factors for deeply virtual Compton
amplitude in the domain\cite{Diehl98,Diehl:1999tr,Muller:1994fv}
$\zeta\le x_1\le 1$:\\
\begin{eqnarray}
&&{\sqrt{1-\zeta}}f_{1\, (n\to n)}(x_1,t,\zeta)\,
-\, {\zeta^2\over 4{\sqrt{1-\zeta}}} f_{2\, (n\to n)}(x_1,t,\zeta)
\nonumber\\
 &=&
\sum_{n, ~ \lambda}
\prod_{i=1}^{n}
\int^1_0 dx_{i(i\ne 1)} \int {d^2{\vec{k}}_{\perp i} \over 2 (2\pi)^3 }
~ \delta\left(1-\sum_{j=1}^n x_j\right) ~ \delta^{(2)}
\left(\sum_{j=1}^n {\vec{k}}_{\perp j}\right)  \nonumber\\[1ex]
&&\times
\psi^{\uparrow \ *}_{(n)}(x^\prime_i, {\vec{k}}^\prime_{\perp i},\lambda_i)
~ \psi^{\uparrow}_{(n)}(x_i, {\vec{k}}_{\perp i},\lambda_i)
(\sqrt{1-\zeta})^{1-n},
\label{t1}
\end{eqnarray}
\begin{eqnarray}
&&
{\sqrt{1-\zeta}}\,\left(\, 1+{\zeta\over 2(1-\zeta)}\,\right)\,
{(\Delta^1-{\mathrm i} \Delta^2)\over 2M}f_{2\, (n\to n)}(x_1,t,\zeta)
\nonumber\\
 &=&
\sum_{n, ~ \lambda}
\prod_{i=1}^{n}
\int^1_0 dx_{i(i\ne 1)} \int {d^2{\vec{k}}_{\perp i} \over 2 (2\pi)^3 }
~ \delta\left(1-\sum_{j=1}^n x_j\right) ~ \delta^{(2)}
\left(\sum_{j=1}^n {\vec{k}}_{\perp j}\right)  \nonumber\\[1ex]
&&\qquad\qquad\qquad\times
\psi^{\uparrow \ *}_{(n)}(x^\prime_i, {\vec{k}}^\prime_{\perp i},\lambda_i)
~ \psi^{\downarrow}_{(n)}(x_i, {\vec{k}}_{\perp i},\lambda_i)
(\sqrt{1-\zeta})^{1-n} ,
\label{t1f2}
\end{eqnarray}
where
\begin{equation}
\left\{ \begin{array}{lll}
x^\prime_1 = {x_1-\zeta \over 1-\zeta}\, ,\
&{\vec{k}}^\prime_{\perp 1} ={\vec{k}}_{\perp 1}
- {1-x_1\over 1-\zeta} {\vec{\Delta}}_\perp
&\mbox{for the struck quark,}\\[1ex]
x^\prime_i = {x_i\over 1-\zeta}\, ,\
&{\vec{k}}^\prime_{\perp i} ={\vec{k}}_{\perp i}
+ {x_i\over 1-\zeta} {\vec{\Delta}}_\perp
&\mbox{for the $ (n-1)$ spectators.}
\end{array}\right.
\label{t2}
\end{equation}
A sum over all possible helicities $\lambda_i$ is understood.
If quark masses are neglected, the currents conserve
helicity.
We also can check that $\sum_{i=1}^n x^\prime_i = 1$,
$\sum_{i=1}^n {\vec{k}}^\prime_{\perp i} = {\vec{0}}_\perp$.

For the $n+1 \to n-1$ off-diagonal term ($\Delta n = -2$),
let us consider the case where
partons $1$ and
$n+1$ of the initial wavefunction annihilate into the current leaving
$n-1$ spectators.
Then $x_{n+1} = \zeta - x_{1}$,
${\vec{k}}_{\perp n+1} = {\vec{\Delta}}_\perp-{\vec{k}}_{\perp 1}$.
The remaining $n-1$ partons have total momentum
$((1-\zeta)P^+, -{\vec{\Delta}}_{\perp})$.
The final wavefunction then has arguments
$x^\prime_i = {x_i \over 1- \zeta}$ and
${\vec{k}}^\prime_{\perp i}=
{\vec{k}}_{\perp i} + {x_i\over 1-\zeta} {\vec{\Delta}}_\perp$.
We thus obtain the formulae for the off-diagonal matrix element
of the Compton amplitude in the domain $0\le x_1\le \zeta$:\\
\begin{eqnarray}
&&{\sqrt{1-\zeta}}f_{1\, (n+1\to n-1)}(x_1,t,\zeta)\,
-\, {\zeta^2\over 4{\sqrt{1-\zeta}}} f_{2\, (n+1\to n-1)}(x_1,t,\zeta)
\nonumber\\
 &=&
\sum_{n, ~ \lambda}
\int^1_0 dx_{n+1}
\int {d^2{\vec{k}}_{\perp 1} \over 2 (2\pi)^3 }
\int {d^2{\vec{k}}_{\perp n+1} \over 2 (2\pi)^3 }
\prod_{i=2}^{n}
\int^1_0 dx_{i} \int {d^2{\vec{k}}_{\perp i} \over 2 (2\pi)^3 }
\nonumber\\[2ex]
&&\times \delta\left(1-\sum_{j=1}^{n+1} x_j\right) ~
\delta^{(2)}\left(\sum_{j=1}^{n+1} {\vec{k}}_{\perp j}\right)
[\sqrt{1-\zeta}]^{1-n}
\nonumber\\[2ex]
&&\times\psi^{\uparrow\ *}_{(n-1)}
(x^\prime_i,{\vec{k}}^\prime_{\perp i},\lambda_i)
~ \psi^{\uparrow}_{(n+1)}(\{x_1, x_i, x_{n+1} = \zeta - x_{1}\},
\nonumber\\[2ex]
&&\qquad\ \ \{ {\vec{k}}_{\perp 1},
{\vec{k}}_{\perp i},
{\vec{k}}_{\perp n+1} = {\vec{\Delta}}_\perp-{\vec{k}}_{\perp 1}\},
\{\lambda_1,\lambda_{i},\lambda_{n+1} = - \lambda_{1}\})
,
\end{eqnarray}
\begin{eqnarray}
&&
{\sqrt{1-\zeta}}\,\Big(\, 1+{\zeta\over 2(1-\zeta)}\,\Big)\,
{(\Delta^1-{\mathrm i} \Delta^2)\over 2M}f_{2\, (n+1\to n-1)}(x_1,t,\zeta)
\nonumber\\
 &=&
\sum_{n, ~ \lambda}
\int^1_0 dx_{n+1}
\int {d^2{\vec{k}}_{\perp 1} \over 2 (2\pi)^3 }
\int {d^2{\vec{k}}_{\perp n+1} \over 2 (2\pi)^3 }
\prod_{i=2}^{n}
\int^1_0 dx_{i} \int {d^2{\vec{k}}_{\perp i} \over 2 (2\pi)^3 }
\nonumber\\[2ex]
&&\qquad\qquad\qquad\times \delta\left(1-\sum_{j=1}^{n+1} x_j\right) ~
\delta^{(2)}\left(\sum_{j=1}^{n+1} {\vec{k}}_{\perp j}\right)
[\sqrt{1-\zeta}]^{1-n}
\nonumber\\[2ex]
&&\qquad\qquad\qquad\times\psi^{\uparrow\ *}_{(n-1)}
(x^\prime_i,{\vec{k}}^\prime_{\perp i},\lambda_i)
~ \psi^{\downarrow}_{(n+1)}(\{x_1, x_i, x_{n+1} = \zeta - x_{1}\},
\nonumber\\[2ex]
&&\qquad\qquad\ \ \{ {\vec{k}}_{\perp 1},
{\vec{k}}_{\perp i},
{\vec{k}}_{\perp n+1} = {\vec{\Delta}}_\perp-{\vec{k}}_{\perp 1}\},
\{\lambda_1,\lambda_{i},\lambda_{n+1} = - \lambda_{1}\})
,
\end{eqnarray}
where $i=2,3,\cdots ,n$
label the $n-1$ spectator
partons which appear in the final-state hadron wavefunction
with
\begin{equation}
x^\prime_i = {x_i\over 1-\zeta}\, ,\qquad
{\vec{k}}^\prime_{\perp i} ={\vec{k}}_{\perp i}
+ {x_i\over 1-\zeta} {\vec{\Delta}}_\perp \ .
\end{equation}
We can again check that the arguments of the final-state wavefunction
satisfy
$\sum_{i=2}^n x^\prime_i = 1$,
$\sum_{i=2}^n {\vec{k}}^\prime_{\perp i} = {\vec{0}}_\perp$.

The above representation is the general form for the generalized form
factors of the deeply virtual Compton amplitude for any composite system.
Thus given the light-cone Fock state wavefunctions of the eigensolutions
of the light-cone Hamiltonian, we can compute the amplitude for virtual
Compton scattering including all spin correlations.  The formulae are
accurate to leading order in
$1/Q^2$.  Radiative corrections to the quark Compton amplitude of order
$\alpha_s(Q^2)$ from diagrams in which a hard gluon interacts between
the two photons have also been neglected.

\section{Electroweak Matrix Elements and Light-Cone Wavefunctions}

\vspace{.5cm}
\begin{figure}[htb]
\begin{center}
\leavevmode
\epsfbox{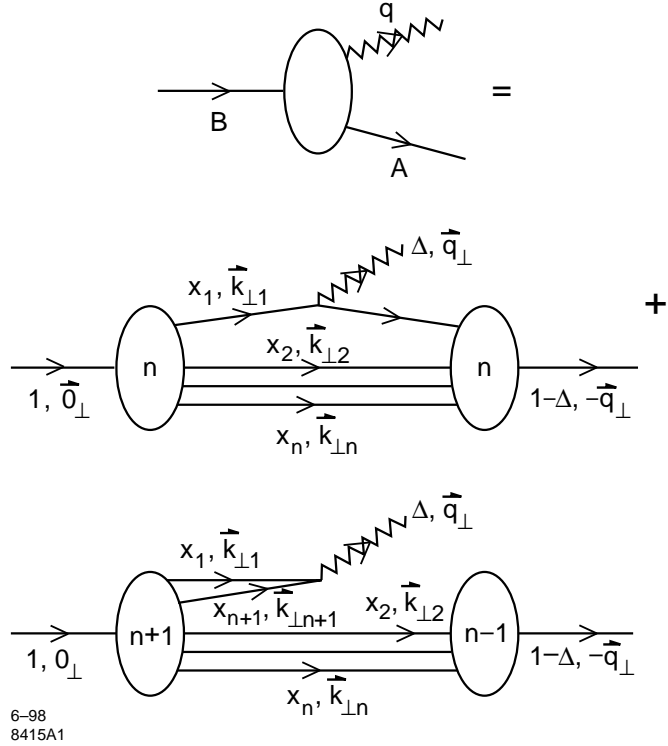}
\end{center}
\caption[*]{Exact representation of electroweak decays and time-like form
factors in the
light-cone Fock representation.
}
\label{fig1}
\end{figure}

Another remarkable advantage of the light-cone formalism is that
exclusive semileptonic
$B$-decay amplitudes such as $B\rightarrow A \ell \bar{\nu}$ can be
evaluated exactly.\cite{Brodsky:1998hn}
The time-like decay matrix elements require the computation of the
diagonal matrix element $n \rightarrow n$ where parton number is
conserved, and the off-diagonal $n+1\rightarrow n-1$ convolution where
the current operator annihilates a $q{\bar{q'}}$ pair in the initial $B$
wavefunction.  See Fig.  \ref{fig1}.  This term is a consequence of the
fact that the time-like decay $q^2 = (p_\ell + p_{\bar{\nu}} )^2 > 0$
requires a positive light-cone momentum fraction $q^+ > 0$.  Conversely
for space-like currents, one can choose $q^+=0$, as in the
Drell-Yan-West representation of the space-like electromagnetic form
factors.  However, as can be seen from the explicit analysis of the form
factor in a perturbative model, the off-diagonal convolution can yield a
nonzero $q^+/q^+$ limiting form as $q^+ \rightarrow 0$.  This extra term
appears specifically in the case of ``bad" currents such as $J^-$ in
which the coupling to $q\bar q$ fluctuations in the light-cone
wavefunctions are favored.  In effect, the $q^+ \rightarrow 0$ limit
generates $\delta(x)$ contributions as residues of the $n+1\rightarrow
n-1$ contributions.  The necessity for such ``zero mode" $\delta(x)$ terms
has been noted by Chang, Root and Yan \cite{CRY}, Burkardt \cite{BUR},
and Ji and Choi.\cite{Choi:1998nf}

The off-diagonal $n+1 \rightarrow n-1$ contributions give a new
perspective for the physics of $B$-decays.  A semileptonic decay
involves not only matrix elements where a quark changes flavor, but also
a contribution where the leptonic pair is created from the annihilation
of a $q {\bar{q'}}$ pair within the Fock states of the initial $B$
wavefunction.  The semileptonic decay thus can occur from the
annihilation of a nonvalence quark-antiquark pair in the initial hadron.
This feature will carry over to exclusive hadronic $B$-decays, such as
$B^0 \rightarrow \pi^-D^+$.  In this case the pion can be produced from
the coalescence of a $d\bar u$ pair emerging from the initial higher
particle number Fock wavefunction of the $B$.  The $D$ meson is then
formed from the remaining quarks after the internal exchange of a $W$
boson.

In principle, a precise evaluation of the hadronic matrix elements
needed for $B$-decays and other exclusive electroweak decay amplitudes
requires knowledge of all of the light-cone Fock wavefunctions of the
initial and final state hadrons.  In the case of model gauge theories
such as QCD(1+1) \cite{Horn} or collinear QCD \cite{AD} in one-space and
one-time dimensions, the complete evaluation of the light-cone
wavefunction is possible for each baryon or meson bound-state using the
DLCQ method.  It would be interesting to use such solutions as a model
for physical $B$-decays.

\section{Applications of Light-Cone Factorization to Hard QCD
 Processes}

Factorization theorems for hard exclusive, semi-exclusive, and
diffractive processes allow a rigorous separation of soft
non-perturbative dynamics of the bound state hadrons from the hard
dynamics of a perturbatively-calculable quark-gluon scattering
amplitude.

The light-cone
formalism provides a physical factorization scheme which conveniently
separates and factorizes soft non-perturbative physics from hard
perturbative dynamics in both exclusive and
inclusive reactions.\cite{Lepage:1980fj,Lepage:1979zb} In hard inclusive
reactions all intermediate states are divided according to $\M^2_n <
\Lambda^2 $ and
$\M^2_n >
\Lambda^2 $ domains.  The lower mass regime is associated with the quark
and gluon distributions defined from the absolute squares of the LC
wavefunctions in the light cone factorization scheme.  In the high
invariant mass regime, intrinsic transverse momenta can be ignored, so
that the structure of the process at leading power has the form of hard
scattering on collinear quark and gluon constituents, as in the parton
model.  The attachment of gluons from the LC wavefunction to a propagator
in a hard subprocess is power-law suppressed in LC gauge, so that the
minimal quark-gluon particle-number subprocesses dominate.  It is then
straightforward to derive the DGLAP equations from the evolution of the
distributions with $\log \Lambda^2$.
The
anomaly contribution to singlet helicity structure function $g_1(x,Q)$
can be explicitly identified in the LC factorization scheme as due to the
$\gamma^* g \to q
\bar q$ fusion process.  The anomaly contribution would be zero if the
gluon is on shell.  However, if the off-shellness of the state is larger
than the quark pair mass, one obtains the usual anomaly
contribution.\cite{Bass:1998rn}

In exclusive amplitudes, the LC wavefunctions are the interpolating
functions between the quark and gluon states and the hadronic states.
In an
exclusive amplitude involving a hard scale $Q^2$ all intermediate states
can be divided according to
$\M^2_n <
\Lambda^2 < Q^2 $ and $\M^2_n < \Lambda^2 $ invariant mass domains.  The
high invariant mass contributions to the amplitude has the structure of a
hard scattering process
$T_H$ in which the hadrons are replaced by their respective (collinear)
quarks and gluons.  In light-cone gauge only the minimal Fock states
contribute to the leading power-law fall-off of the exclusive amplitude.
The wavefunctions in the lower invariant mass domain can be integrated up
to the invariant mass cutoff $\Lambda$.  Final-state
and initial state corrections from gluon attachments to lines
connected to the color-singlet distribution amplitudes cancel at
leading twist.

Given the solution
for the hadronic wavefunctions $\psi^{(\Lambda)}_n$ with $\M^2_n <
\Lambda^2$, one can construct the wavefunction in the hard regime with
$\M^2_n > \Lambda^2$ using projection operator techniques.\cite{LB}  The
construction can be done perturbatively in QCD since only high invariant mass,
far off-shell matrix elements are involved.  One can use this method to
derive the physical properties of the LC wavefunctions and their matrix elements
at high invariant mass.  Since $\M^2_n = \sum^n_{i=1}
\left(\frac{k^2_\perp+m^2}{x}\right)_i $, this method also allows the derivation
of the asymptotic behavior of light-cone wavefunctions at large $k_\perp$, which
in turn leads to predictions for the fall-off of form factors and other
exclusive
matrix elements at large momentum transfer, such as the quark counting rules
for predicting the nominal power-law fall-off of two-body scattering amplitudes
at fixed
$\theta_{cm}$.\cite{BrodskyLepage}  The phenomenological successes of
these rules
can be understood within QCD if the coupling
$\alpha_V(Q)$ freezes in a range of relatively
small momentum transfer.\cite{BJPR}

The key non-perturbative input for exclusive
processes is the gauge and frame independent hadron distribution
amplitude \cite{Lepage:1979zb,Lepage:1980fj} defined as the integral of
the valence (lowest particle number) Fock wavefunction;
\eg\ for the pion
\begin{equation}
\phi_\pi (x_i,\Lambda) \equiv \int d^2k_\perp\, \psi^{(\Lambda)}_{q\bar
q/\pi} (x_i, \vec k_{\perp i},\lambda)
\label{eq:f1a}
\end{equation}
where the global cutoff $\Lambda$ is identified with the resolution $Q$.
The distribution amplitude controls leading-twist exclusive amplitudes
at high momentum transfer, and it can be related to the gauge-invariant
Bethe-Salpeter wavefunction at equal light-cone time.  The
logarithmic evolution of hadron distribution amplitudes
$\phi_H (x_i,Q)$ can be derived from the perturbatively-computable tail
of the valence light-cone wavefunction in the high transverse momentum
regime.\cite{Lepage:1979zb,Lepage:1980fj} The conformal basis for the
evolution of the three-quark distribution amplitudes for
the baryons~\cite{Lepage:1979za} has recently been obtained by V. Braun
\etal\cite{Braun:1999te}

Thus at high transverse
momentum an exclusive amplitudes factorizes into a convolution of a hard
quark-gluon subprocess amplitudes $T_H$ with the hadron distribution
amplitudes
$\phi(x_i,\Lambda)$.\cite{BL80} The $T_H$ satisfy the dimensional
counting rules.  The logarithmic evolution of hadron distribution
amplitudes
$\phi_H (x_i,Q)$ can be derived from the perturbatively-computable tail
of the valence light-cone wavefunction in the high transverse momentum
regime.\cite{LB}

The existence of an exact formalism
provides a basis for systematic approximations and a control over neglected
terms.  For example, one can analyze exclusive semi-leptonic
$B$-decays which involve hard internal momentum transfer using a
perturbative QCD formalism\cite{BHS,Beneke:1999br}
patterned after the analysis
of form
factors at large momentum transfer.\cite{LB}  The hard-scattering
analysis again proceeds
by writing each hadronic wavefunction
as a sum of soft and hard contributions
\begin{equation}
\psi_n = \psi^{{\rm soft}}_n (\M^2_n < \Lambda^2) + \psi^{{\rm hard}}_n
(\M^2_n >\Lambda^2) ,
\end{equation}
where $\M^2_n $ is the invariant mass of the partons in the $n$-particle
Fock state and
$\Lambda$ is the separation scale.
The high internal momentum contributions to the wavefunction $\psi^{{\rm
hard}}_n $ can be calculated systematically from QCD perturbation theory
by iterating the gluon exchange kernel.  The contributions from
high momentum transfer exchange to the
$B$-decay amplitude can then be written as a
convolution of a hard-scattering
quark-gluon scattering amplitude $T_H$ with the distribution
amplitudes $\phi(x_i,\Lambda)$, the valence wavefunctions obtained by
integrating the
constituent momenta up to the separation scale
${\cal M}_n < \Lambda < Q$.  Furthermore in processes such as $B \to \pi
D$ where the pion is effectively produced as a rapidly-moving small Fock
state with a small color-dipole interactions,  final state interactions
are suppressed by color transparency.  This is the basis for the
perturbative hard-scattering analyses.\cite{BHS,Sz,BABR,Beneke:1999br}
In the exact analysis, one can
identify the hard PQCD contribution as well as the soft contribution from
the convolution of the light-cone wavefunctions.
Furthermore, the hard-scattering contribution
can be systematically improved.

\section{Non-Perturbative Solutions of Light-Cone Quantized QCD}

It is clearly important  not only to compute the spectrum of hadrons and
gluonic states, but also to determine the wavefunction of each QCD bound
state in terms of its fundamental quark and gluon degrees of freedom.
If we could obtain such nonperturbative solutions of QCD, then we could
compute the quark and gluon structure functions and distribution
amplitudes which control hard-scattering inclusive and exclusive
reactions as well as calculate the matrix elements of currents which
underlie electroweak form factors and the weak decay amplitudes of the
light and heavy hadrons.  The light-cone wavefunctions also determine
the multi-parton correlations which control the distribution of
particles in the proton fragmentation region as well as dynamical higher
twist effects.  Thus one can analyze not only the deep inelastic
structure functions but also the fragmentation of the spectator system.
Knowledge of hadron wavefunctions would also open a window to a deeper
understanding of the physics of QCD at the amplitude level, illuminating
exotic effects of the theory such as color transparency, intrinsic heavy
quark effects, hidden color, diffractive processes, and the QCD van der
Waals interactions.

Is there any hope of computing light-cone wavefunctions from
first principles?  In the discretized light-cone quantization
method (DLCQ),\cite{Pauli:1985ps}
periodic boundary conditions are introduced in
$b_\perp$ and $x^-$ so that the momenta
$k_{\perp i} = n_\perp \pi/ L_\perp$ and $x^+_i = n_i/K$ are
discrete.  A global cutoff in invariant mass of the partons in the Fock
expansion is also introduced.
Solving the quantum field theory then reduces to
the problem of diagonalizing the finite-dimensional hermitian matrix
$H_{LC}$ on a finite discrete Fock basis.  The DLCQ method has now become
a standard tool for solving both the spectrum and light-cone wavefunctions
of one-space one-time theories -- virtually any
$1+1$ quantum field theory, including ``reduced QCD" (which has both quark and
gluonic degrees of freedom) can be completely solved using
DLCQ.\cite{Dalley:1993yy,Kleb,AD}
The method yields not only the bound-state and continuum
spectrum, but also the light-cone wavefunction for each eigensolution.  The
solutions for the model 1+1 theories can provide an important theoretical
laboratory for testing approximations and QCD-based models.

In the case of theories in 3+1 dimensions, Hiller, McCartor, and I\
\cite{Brodsky:1998hs,Brodsky:1999xj} have recently shown that the use of
covariant Pauli-Villars regularization with DLCQ allows one to obtain the
spectrum and light-cone wavefunctions of simplified theories, such as
(3+1) Yukawa theory.  Dalley \etal\ have shown how one can use DLCQ
in one space-one time, with a transverse lattice to solve mesonic
and gluonic states in $ 3+1$ QCD.\cite{Dalley:1999ii} The spectrum
obtained for gluonium states
is in remarkable agreement with lattice gauge theory results, but with a
huge reduction of numerical effort.  Hiller and I \cite{Hiller:1999cv}
have shown how one can use DLCQ to compute the electron magnetic moment
in QED without resort to perturbation theory.   Light-cone gauge $A^+ =
0$ allows one to utilize
only the physical degrees of freedom of the gluon field to appear.
However, light-cone quantization in Feynman gauge has a number of
attractive features, including manifest covariance and a straightforward
passage to the Coulomb limit in the case of static
quarks.\cite{Srivastava:1999gi}

One can also formulate DLCQ so
that supersymmetry is exactly preserved in the discrete approximation,
thus combining the power of DLCQ with the beauty of
supersymmetry.\cite{Antonuccio:1999ia,Lunin:1999ib,Haney:1999tk} The
``SDLCQ" method has been applied to several interesting supersymmetric
theories, to the analysis of zero modes, vacuum degeneracy, massless
states, mass gaps, and theories in higher dimensions, and even tests of
the Maldacena conjecture.\cite{Antonuccio:1999ia}

Broken supersymmetry is
interesting in DLCQ, since it may serve as a method for regulating
non-Abelian theories.\cite{Brodsky:1999xj}
Another
remarkable advantage of light-cone quantization is that the vacuum state
$\ket{0}$ of the full QCD Hamiltonian coincides with the free vacuum.
For example, as discussed by Bassetto,\cite{Bassetto:1999tm}
the computation of the spectrum
of $QCD(1+1)$ in equal time quantization requires constructing the full
spectrum of non perturbative contributions (instantons).  However,
light-cone methods with infrared regularization give the correct result
without any need for vacuum-related contributions.  The role
of instantons and such phenomena in light-cone quantized
$QCD(3+1)$ is presumably more complicated and may reside in zero
modes; \cite{Yamawaki:1998cy}
\eg, zero modes are evidently necessary to represent theories with
spontaneous symmetry breaking.\cite{Pinsky:1994yi}

Light-cone wavefunctions thus are the natural quantities to encode hadron
properties and to bridge the gap between empirical constraints and
theoretical predictions for the bound state solutions.  We can thus
envision a program to construct the hadronic light cone Fock
wavefunctions $\psi_n(x_i, k_{\perp i},
\lambda_i)$ using not only data but constraints such as:

(1) Since the state is far off shell at large invariant mass $\M$,
one can derive rigorous limits on the
$x \to 1$, high $k_\perp$, and high
$\M^2_n$ behavior of the wavefunctions in the perturbative domain.

(2) Ladder relations connecting state of different particle number
follow from the QCD equation of motion and lead to Regge behavior of the
quark and gluon distributions at $x \to 0$.  QED provides a constraint at
$N_C \to 0.$

(3) One can obtain guides to the exact behavior of LC wavefunctions in
QCD from analytic or DLCQ solutions to toy models such as ``reduced"
$QCD(1+1).$

(4) QCD sum rules, lattice gauge theory moments, and QCD inspired models
such as the bag model, chiral theories, provide important constraints.
An important question is how the light-cone wavefunctions incorporate
chiral constraints such as soliton (Skyrmion) behavior for baryons and
other consequences of the chiral limit.  However it has been shown that
the anomaly contribution for the $\pi^0\to \gamma \gamma$ decay amplitude
is satisfied by the light-cone Fock formalism in the limit where the mass
of the pion is light compared to its size.\cite{Lepage:1982gd}

(5) Since the LC formalism is valid at all scales, one can utilize
empirical constraints such as the measurements of magnetic
moments, axial couplings, form factors, and distribution amplitudes.

(6) In the nonrelativistic limit, the light-cone and many-body
Schr\"odinger theory formalisms must match.

\section{Self-Resolved Diffractive Reactions and Light Cone Wavefunctions}

Diffractive multi-jet production in heavy
nuclei provides a novel way to measure the shape of the LC Fock
state wavefunctions and test color transparency.  For example, consider the
reaction \cite{Bertsch,MillerFrankfurtStrikman,Frankfurt:1999tq}
$\pi A \rightarrow {\rm Jet}_1 + {\rm Jet}_2 + A^\prime$
at high energy where the nucleus $A^\prime$ is left intact in its ground
state.  The transverse momenta of the jets balance so that
$
\vec k_{\perp i} + \vec k_{\perp 2} = \vec q_\perp < {R^{-1}}_A \ .
$
The light-cone longitudinal momentum fractions also need to add to
$x_1+x_2 \sim 1$ so that $\Delta p_L < R^{-1}_A$.  The process can
then occur coherently in the nucleus.  Because of color transparency, the
valence wavefunction of the pion with small impact separation, will
penetrate the nucleus with minimal interactions, diffracting into jet
pairs.\cite{Bertsch} The $x_1=x$, $x_2=1-x$ dependence of
the di-jet distributions will thus reflect the shape of the pion
valence light-cone wavefunction in $x$; similarly, the
$\vec k_{\perp 1}- \vec k_{\perp 2}$ relative transverse momenta of the
jets gives key information on the derivative of the underlying shape
of the valence pion
wavefunction.\cite{MillerFrankfurtStrikman,Frankfurt:1999tq,BHDP} The
diffractive nuclear amplitude extrapolated to
$t = 0$ should be linear in nuclear number $A$ if color transparency is
correct.  The integrated diffractive rate should then scale as $A^2/R^2_A
\sim A^{4/3}$.  Preliminary results on a diffractive dissociation
experiment of this type E791 at Fermilab using 500 GeV incident pions on
nuclear targets.\cite{Ashery:1999nq} appear to be consistent with color
transparency.\cite{Ashery:1999nq} The momentum fraction distribution of
the jets is consistent with a valence light-cone wavefunction of the pion
consistent with the shape of the asymptotic distribution amplitude,
$\phi^{\rm asympt}_\pi (x) =
\sqrt 3 f_\pi x(1-x)$.  Data from CLEO \cite{Gronberg:1998fj} for the
$\gamma \gamma^* \rightarrow \pi^0$ transition form factor also favor a form for
the pion distribution amplitude close to the asymptotic solution
\cite{Lepage:1979zb,Lepage:1980fj} to the perturbative QCD evolution
equation.

The diffractive dissociation of a hadron or nucleus can also occur via
the Coulomb dissociation of a beam particle on an electron beam (\eg\ at
HERA or eRHIC) or on the strong Coulomb field of a heavy nucleus (\eg\
at RHIC or nuclear collisions at the LHC).\cite{BHDP} The amplitude for
Coulomb exchange at small momentum transfer is proportional to the first
derivative $\sum_i e_i {\partial \over \vec k_{T i}} \psi$ of the
light-cone wavefunction, summed over the charged constituents.  The Coulomb
exchange reactions fall off less fast at high transverse momentum compared
to pomeron exchange reactions since the light-cone wavefunction is
effective differentiated twice in two-gluon exchange reactions.

It will also be interesting to study diffractive tri-jet production
using proton beams
$ p A \rightarrow {\rm Jet}_1 + {\rm Jet}_2 + {\rm Jet}_3 + A^\prime $ to
determine the fundamental shape of the 3-quark structure of the valence
light-cone wavefunction of the nucleon at small transverse
separation.\cite{MillerFrankfurtStrikman}
For example, consider the Coulomb dissociation of a high energy proton at
HERA.  The proton can dissociate into three jets corresponding to the
three-quark structure of the valence light-cone wavefunction.  We can
demand that the produced hadrons all fall outside an opening angle $\theta$
in the proton's fragmentation region.
Effectively all of the light-cone momentum
$\sum_j x_j \simeq 1$ of the proton's fragments will thus be
produced outside an ``exclusion cone".  This
then limits the invariant mass of the contributing Fock state ${\cal
M}^2_n >
\Lambda^2 = P^{+2} \sin^2\theta/4$ from below, so that perturbative QCD
counting rules can predict the fall-off in the jet system invariant mass
$\cal M$.  At large invariant mass one expects the three-quark valence
Fock state of the proton to dominate.  The segmentation of the forward
detector in azimuthal angle $\phi$ can be used to identify structure and
correlations associated with the three-quark light-cone
wavefunction.\cite{BHDP}
An interesting possibility is
that the distribution amplitude of the
$\Delta(1232)$ for $J_z = 1/2, 3/2$ is close to the asymptotic form $x_1
x_2 x_3$,  but that the proton distribution amplitude is more complex.
This ansatz can also be motivated by assuming a quark-diquark structure
of the baryon wavefunctions.  The differences in shapes of the
distribution amplitudes could explain why the $p
\to\Delta$ transition form factor appears to fall faster at large $Q^2$
than the elastic $p \to p$ and the other $p \to N^*$ transition form
factors.\cite{Stoler:1999nj} One can use also measure the dijet
structure of real and virtual photons beams
$ \gamma^* A \rightarrow {\rm Jet}_1 + {\rm Jet}_2 + A^\prime $ to
measure the shape of the light-cone wavefunction for
transversely-polarized and longitudinally-polarized virtual photons.  Such
experiments will open up a direct window on the amplitude
structure of hadrons at short distances.
The light-cone formalism is also applicable to the
description of nuclei in terms of their nucleonic and mesonic
degrees of freedom.\cite{Miller:1999mi,Miller:2000ta}
Self-resolving diffractive jet reactions
in high energy electron-nucleus collisions and hadron-nucleus collisions
at moderate momentum transfers can thus be used to resolve the light-cone
wavefunctions of nuclei.

\section{Dynamical Correlations and Higher-Twist \hfill\break Effects in QCD}

It is an empirical fact that conventional leading twist contributions
cannot account for the measured $e p
\rightarrow e X$ and $e d \rightarrow e X$ structure functions at $x
\gsim$ 0.4 and $Q^2 \lsim$ 5 GeV$^2$.  Fits to the
data \cite{Virchaux:1992jc,Amaudruz:1992nw} require an additional
component which scales as
$1/Q^2$ relative to the leading twist contributions and rises
rapidly with
$x$.  The excess contribution can be parameterized in the form
\begin{figure}[htb]
\begin{center}
\leavevmode
{\epsfxsize=3in\epsfbox{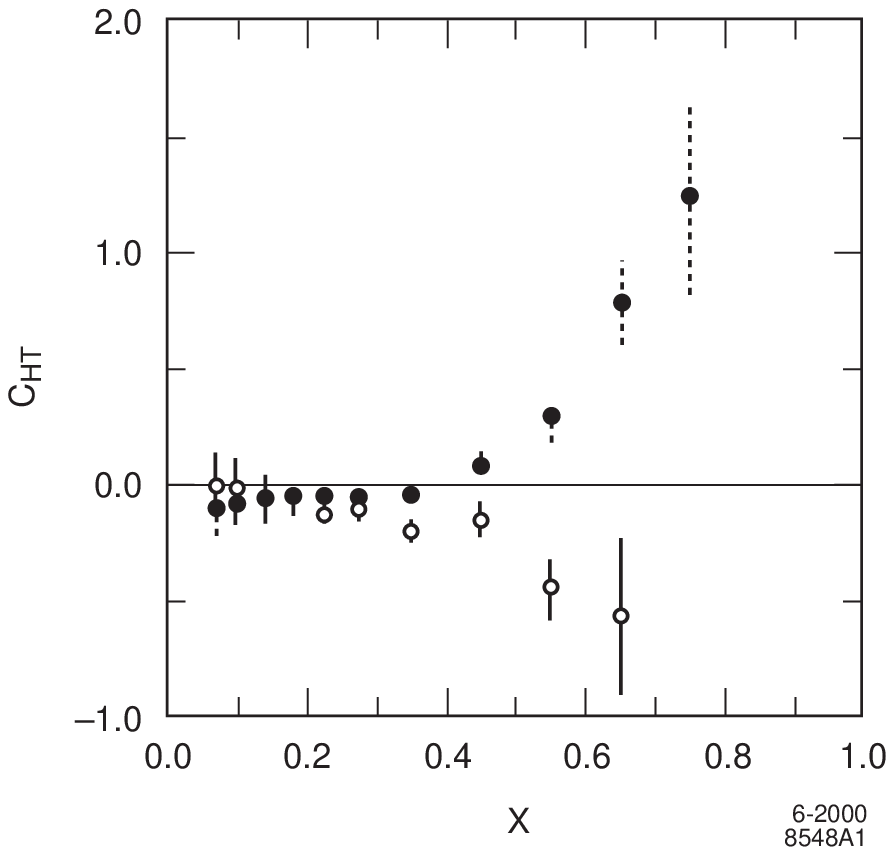}}
\caption[*]{Higher-twist coefficients $C_{HT}(x)$ [in
GeV$^2$ units] for
inelastic lepton scattering on proton target (solid points) and the
difference $C^p_{HT}(x)-C^n_{HT}(x)$ for proton minus neutron targets
(open circles).\cite{Virchaux:1992jc,Amaudruz:1992nw}  The data compilation
is taken from Souder.\cite {Souder:2000}}
\label{fig:8548A1}
\end{center}
\end{figure}

\begin{equation}
F_{2p,n}(x,Q^2) = F^0_{2p,n}(x,Q^2)
\left[1+ \frac{c_{HT}^{p,n}(x)}{Q^2}\right]
\label{eq:al}
\end{equation}
where $F^0_{2p,n}$ is the leading twist contribution.  The functional
dependence of the higher-twist term $C_{HT}^{p,n}(x)$ for proton and
proton-neutron targets is shown in Fig. \ref{fig:8548A1}.  A rough fit is
\begin{equation}
c_{HT}^p(x) \cong \left[\frac{0.3\ GeV}{1-x}\right]^2
\qquad
c_{HT}^n(x) \cong 2 c_{HT}^p(x) \ ;
\label{eq:am}
\end{equation}
{\it i.e.}:
the higher-twist effect relative to the leading twist contribution for
the neutron is stronger than that of the proton.

A possible source of higher-twist effects in PQCD
is ``renormalons''.\cite{Beneke:1999ui,Maul:1997rz}  This
contribution to the deep inelastic lepton-hadron cross section reflects
a divergent $\beta^n_0\, n$! growth of the PQCD series for hard
radiative corrections to deep inelastic scattering evolution at
high orders in $\alpha^n_s(Q^2)$.  The factorial growth arises from the
integration over the QCD running coupling; {\it i.e.}, the summation
of the reducible multi-bubble loop-diagrams in the gluon propagator.
The net
effect is to correct the leading twist predictions by a power-law
suppressed $1/Q^2(1-x)$ contribution.
Alternatively, one can proceed using the BLM method \cite{Brodsky:1983gc}:
one first
identifies the conformal coefficients\cite{Brodsky:2000cr} of the
PQCD series; by definition these are independent of the
$\beta-$function and are hence devoid of the
$\beta^n_0\, n!$ growth.  The scale of the running coupling is
set by requiring that all of the
$\beta$-dependence resides in $\alpha_s(Q^{*2}).$
The resulting scale
$(Q^{*2}) \propto (1-x)Q^2$ can also be understood as the mean value
of the argument of the running coupling $\alpha_s(k^2)$ in the Feynman
loop integration.

However, the renormalon contribution cannot account
for the observed higher-twist contribution shown in Fig. \ref{fig:8548A1}
since it is proportional to the leading-twist prediction, {\it i.e.}:
$c_{HT ren}^p(x) = c_{HT ren }^n(x) .$
Thus it is apparent from the data that there must be a dynamical origin
for the observed $C_{HT}(x)/Q^2$ contribution.  In fact, dynamical
higher-twist terms naturally arise from multi-parton correlations.  For
example, if the electron recoils against 1, 2, or 3 quarks, one obtains
a series of higher-twist contributions of ascending order in $1/Q^2$.
\begin{eqnarray}
\sigma_T &\sim& \frac{(1-x)^3}{Q^2)} \qquad e q \rightarrow eq \nonumber
\\[1ex]
\sigma_L &\sim& \frac{(1-x)}{(Q^2)^3} \qquad e qq \rightarrow e qq \\[1ex]
\sigma_T &\sim& \frac{1}{(1-x)} \left(\frac{1}{Q^2}\right)^3 \quad e qqq
\rightarrow
e qqq \nonumber
\label{eq:ao}
\end{eqnarray}
where the extra $1/Q^2$ fall-off reflects the form factor squared of the
$(qq)$ or
$(qqq)$ systems, and the enhancement at $x \rightarrow 1$ reflects the
fact that the $(qq)$ and $(qqq)$ composites carry increasing fractions of
the proton light-cone momentum.  The dominance of $\sigma_L$ for $eqq
\rightarrow eqq$ reflects the bosonic coupling of the composite di-quark.
Each of the contributions satisfy Bloom-Gilman
duality \cite{Bloom:1970xb} at fixed $W^2$.  The multi-parton subprocesses
are suppressed by powers of
$1/Q^2$ but enhanced at large $x$ since more of the momentum of the
target proton is fed into the hard subprocess; {\it i.e.}, there are fewer
spectators to stop.  The general rule is
\begin{equation}
F_2(x,Q^2)\propto {(1-x)^{2 n_{spect} - 1 + 2 \Delta h}\over
Q^{n_{active}-4}}
\end{equation}
where $n$ is the number of partons or other quanta
participating in the hard subprocess and $\Delta h$ is the difference in
helicity between the active partons and the
target.\cite{Blankenbecler:1978vk}

\vspace{.5cm}
\begin{figure}[htb]
\begin{center}
\leavevmode
{\epsfbox{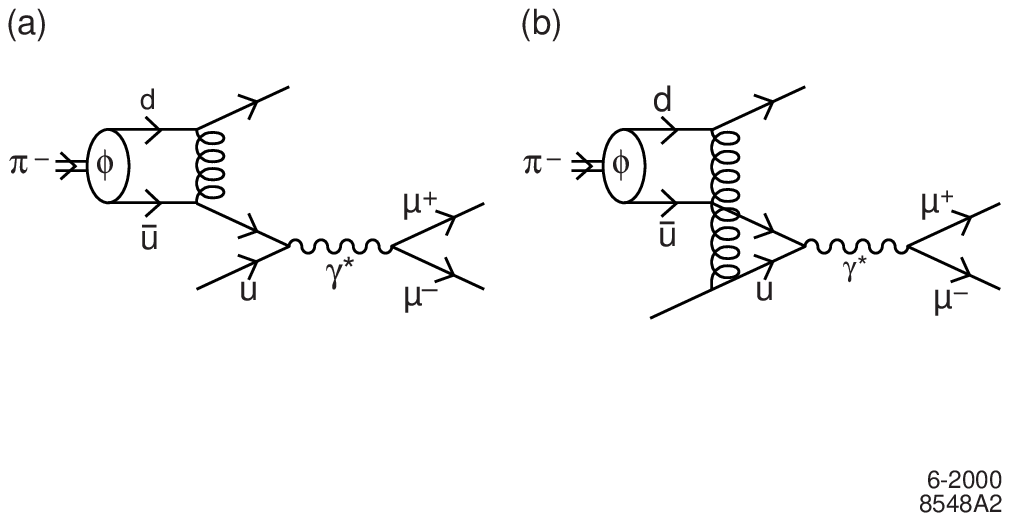}}
\caption[*]{Higher-twist contribution to lepton pair
production in $\pi
N$ scattering.  The dynamics at large $x_F$ requires both constituents of
the projectile meson to be involved in the hard subprocess.
\cite{Brandenburg:1994wf}}
\label{fig:8548A2}
\end{center}
\end{figure}

It is well-known that higher-twist, power-law suppressed corrections to
hard inclusive cross sections can be a signature of correlation
effects involving two or more valence quarks of a hadron.  For example,
the lepton angular dependence of the leading-twist PQCD prediction for
Drell Yan lepton pair production
$d\sigma (\pi A \to \ell^+ \ell^- X)/d \Omega$ is $1 + \cos^2
\theta_{cm}$.  The data\cite{Guanziroli:1988rp,Conway:1989fs}
however shows the onset of $\sin^2 \theta_{cm}$ dependence at large
$x_F$.  This signals the presence of multiparton-induced subprocesses
such as
$(\bar q q) q \to \gamma^*(Q^2) q \to \ell^+ \ell^- q$.\cite{Berger:1979du}
See Fig. \ref{fig:8548A2}.
Such reactions
produce longitudinally-polarized virtual photons with a $\sin^2
\theta_{cm}$ lepton pair angular dependence in contrast to the
transversally polarized Drell-Yan pairs produced from the
${\bar q } q \to \gamma^*(Q^2) \to \ell^+ \ell^-$ subprocess.
The penalty for utilizing the two correlated partons in the pion
wavefunction is an extra suppression factor $1/ R^2 Q^2 (1-x_F)^2$
where
$R$ is the characteristic interquark transverse separation between the
valence quarks in the incoming meson.  The origin of the $1/R^2 Q^2$
scaling is similar to that of the photon to meson transition form factor
in the exclusive reaction $\ell \gamma \to \ell (\bar q q) \to
\pi^0$.\cite{BL80}  The scale $1/R$ can be related to the pion decay
constant $f_\pi$ which normalizes the pion distribution
amplitude.\cite{BL80}   At fixed
$Q^2$ the higher-twist process can actually dominate as $x_F \to 1$ since
all of the incoming momentum of the pion is transferred to the
subprocess.  The correlated subprocess
$(\bar q q) q
\to \gamma^*(Q^2) q \to \ell^+ \ell^- q$ also leads to the prediction of
$\sin^2\theta \cos 2 \phi$ and $\sin 2 \theta \cos \phi$
terms in the angular distribution\cite{Brandenburg:1994wf}, effects which
are clearly apparent in the data.
\cite{Guanziroli:1988rp,Conway:1989fs}

Another important example of dynamical higher-twist effects is the
reaction $\pi A
\to J/\psi X$ which is observed to produce longitudinally-polarized
$J/\psi's$ at large
$x_F$.\cite{Biino:1987qu}  Again this effect can be attributed to highly
correlated multi-parton subprocesses such as $\bar q q g \to \bar c c
\bar q q$ where both valence quarks of the incident pion must be involved
in the hard subprocess in order to produce the charmed quark pair with
nearly all of the incident momentum of the incoming
meson.\cite{Vanttinen:1995sd}  Similarly, charm production at
threshold requires that all of the momentum of the target nucleon
be transferred to the charm quarks.  In the $\gamma p \rightarrow
c\overline{c}p$ reaction near threshold,  all the partons have to
transfer their energy to the charm quarks within their reaction time
$1/m_c$, and must be within this transverse distance from the
$c\overline{c}$ and from each other.
Hence only compact Fock states of the
target nucleon or nucleus with a radius equal to the Compton wavelength of
the heavy quark, can contribute to charm production at threshold.
Equivalently we can interpret the multi-connected charm quarks as
intrinsic charm Fock states which are kinematically favored to have
large momentum fractions.\cite{Brodsky:1980pb}
The experimental evidence for intrinsic charm is discussed
by Harris, Smith, and Vogt.\cite{Harris:1996jx}

Near-threshold charm production also probes the $x\simeq 1$ configurations
in the target wavefunction; the spectator partons carry a vanishing
fraction
$x\simeq 0$ of the target momentum.  This implies that the production
rate behaves near $x\rightarrow 1$ approximately as $(1-x)^{2n_s-1}$
where
$n_s$ is the number of spectators required to stop.  Including spin
factors, we can identify three different gluonic components of the
photoproduction cross-section:
\begin{itemize}
\item The usual one-gluon $(1-x)^4$ distribution for leading twist
photon-gluon fusion $\gamma g\rightarrow c \overline{c}$, which leaves
two quarks spectators;
\item Two correlated gluons emitted from the proton with a net
distribution \hfill\break ${(1-x)^2}/{R^2{\cal{M}}^2}$ for $\gamma gg
\rightarrow
c\overline{c}$, leaving one quark spectator;
\item Three correlated gluons emitted from the proton with a net
distribution \hfill\break $
{(1-x)^0}/{R^4{\cal{M}}^4}$ for $\gamma ggg \rightarrow
c\overline{c}$, leaving no quark spectators.
\end{itemize} Here $x \approx {\cal{M}}^2/(s-m^2)$ and $\cal{M}$ is the
mass of the $c\overline{c}$ pair.  The relative weight of the
multiply-connected terms is controlled by the inter-quark separation
$R\simeq 1/m_c$.  The extra powers of $1/\cal{M}$ arise from the
power-law fall-off of the higher-twist hard subprocesses.\cite{BHL}

\begin{figure}[htb]
\begin{center}
\leavevmode
{\epsfbox{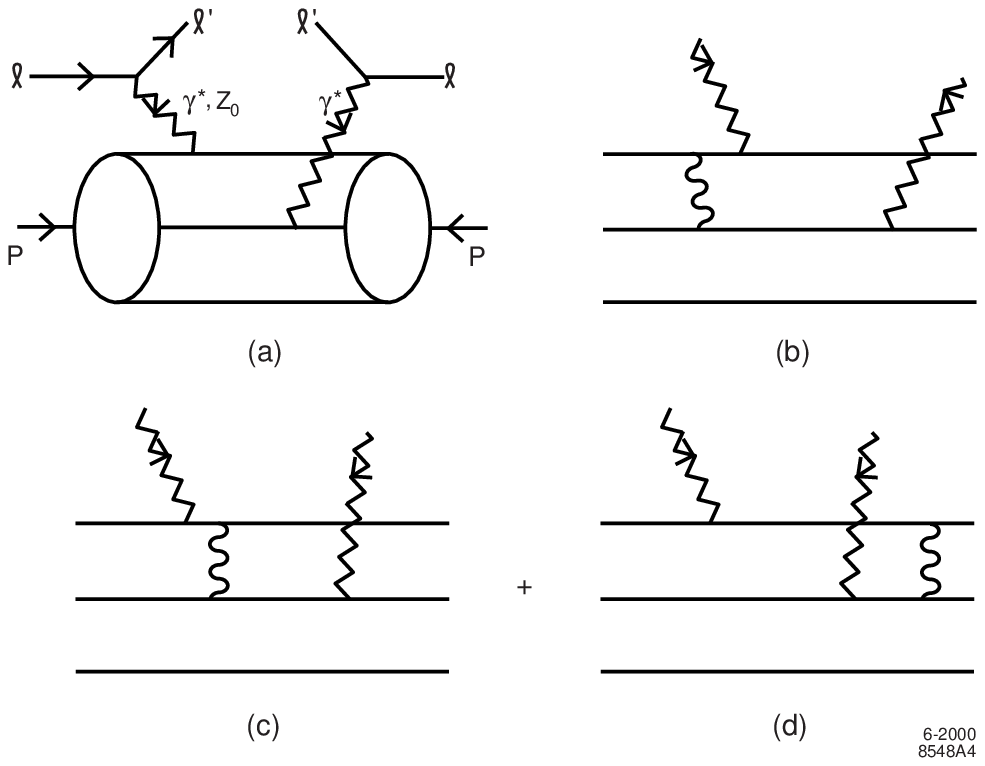}}
\caption[*]{(a) Twist-four contribution to inelastic lepton
scattering where the lepton scatters on different quarks.  The
interference of
$\gamma^*$ and
$Z^0$ exchange contributions leads to parity and charge-conjugation
violation of the higher-twist contribution.  (b-d) The leading-order
${\cal O} (\alpha_s / Q^2 R^2)$ perturbative QCD gluon-exchange
contributions.  The higher-twist contribution to the structure function is
obtained by a convolution of the nucleon light-cone wavefunctions with the
$\gamma^* (q q) \to \gamma^* (q q)$ multi-quark amplitude.}
\label{fig:8548A4}
\end{center}
\end{figure}

The correlations between valence quarks can also have an important effect
in deep inelastic scattering, particularly at large $x_{bj}= {Q^2/ 2 p
\cdot q}$.   As noted above,
one expects a sum of contributions to the
deep inelastic cross section scaling nominally as
\begin{equation}
F_2(x,Q^2) = A(1-x)^3 + B{(1-x)^2\over Q^4} + C{(1-x)^{-1}\over Q^8}
\end{equation}
corresponding to the subprocesses $\ell q \to \ell q$, $\ell ( q q) \to
\ell (q q)$, and $\ell ( q qq ) \to \ell (q q q)$.
However, the above classification of terms in $F_2(x,Q^2)$ neglects what
may be the most significant and interesting higher-twist contribution to
deep inelastic scattering: the interference contributions.  Let us
consider the contribution to DIS due to the interference of the amplitude
where the lepton scatters on one quark with the amplitude where the lepton
scatters on another quark.  See Fig. \ref{fig:8548A4}.  One might think
such contributions are assumed to be negligible since the hard
subprocesses seem to lead to different non-interfering final states.
Actually these contributions can interfere if the struck quarks have high
internal momentum in the initial state or if they exchange large momenta
in the final state.  In either case, the apparently different final states
can overlap.  An insightful nuclear physics analog has been discussed by
Drell.\cite{Drell:1992rx}

\vspace{.5cm}

Let us consider the electroproduction subprocess
$\ell (q q) \to
\ell q q$ where the initial
$(q q)$ are collinear and have small invariant mass in the initial state
and the
$q q$ pair in the final state can have large invariant mass.  The lepton
can effectively scatter on either quark.  The nominal scaling of such
twist-four contributions is
\begin{equation}
F^{\rm interference}_2(x,Q^2) = \sum_{a \ne b}
e_a e_b {(1-x)^2\over R_{a b}^2 Q^2 }
\end{equation}
where the factor of $1/R^2_{a b}$ reflects the inter-parton distance.
The interference terms are distinctive since, unlike renormalon
contributions, they do not track with the leading twist contributions.
The growth at high $x$ of the twist-four process reflects the fact that
the
$\ell (q q) \to
\ell q q$ subprocess incorporates the momentum of both quarks.  This
contribution must also play an important role in the physics of
Bloom-Gilman duality
\cite{Bloom:1970xb} since the interference contributions also appear in
square of the transition form factors.  The interference terms can
contribute to both
$F_L$ and $F_T$.  There is an extensive literature on higher-twist
contributions to the structure functions coming from such four-fermion
operators.\cite{Capitani:1999ai,Edelmann:2000yp}  They are also referred
to as ``cat ear" diagrams from their appearance in the virtual Compton
amplitude.

Let us suppose that the proton wavefunction is symmetric in the
coordinates of the three valence quarks.  If we sum over the
pairs of valence quarks, we obtain a vanishing contribution on a proton
target
\begin{equation}
\sum_{a \ne b} e_a e_b = (\sum_a e_a)^2 - \sum_a e_a^2 = 1 -
(4/9+4/9+1/9) = 0.
\end{equation}
However, for the neutron
\begin{equation}
\sum_{a \ne b} e_a e_b = (\sum_a e_a)^2 - \sum_a e_a^2 = 0 -
(4/9+1/9+1/9) = 2/3.
\end{equation}
Thus for symmetric nucleon wavefunctions the dynamical higher-twist
cross terms appear to be are zero in the proton and significant for the
neutron, deuteron, and nuclei!  This is a very distinctive effect; it
particularly motivates the empirical study of higher-twist effects using
the deuteron and nuclear targets.

In a more realistic treatment, one needs to take into account
correlation substructure.  For example, suppose that we
can approximate the nucleon wavefunctions as quark di-quark composites,
where the di-quark has $I=0$ and $J=0.$ Let us also suppose that
the inter-quark separation
$R_{a b}$ is smallest for the two quarks of the diquark composite.  In
this case we can approximate the full sum as a sum over the quark charges
of the $I= 0$ $u d$ diquark.  Then
$\sum_{a \ne b} e_a e_b = e_u e_d = -2/9$
for both the proton and neutron targets.  However, since it is
conventional to parameterize the higher-twist contribution as
a correction to the leading twist term.  Thus $C_{p,n}(x)$
is predicted to rise strongly at large $x$ and
$C_n(x) $ will be larger than $C_p(x)$ since the leading-twist
contribution to the neutron structure function
$F^{n}_2(x,Q^2)$ is significantly smaller than $F^{p}_2(x,Q^2)$.  These
predictions seem consistent with the empirical higher-twist contributions
to electroproduction extracted in the
references.\cite{Virchaux:1992jc,Amaudruz:1992nw}
A simple test of the $I = 0$ diquark
higher-twist model  is the absence of twist-four contributions to the
combination of structure functions
$F^{d}_2(x,Q^2) - 2 F^{p}_2(x,Q^2)$

It is also
interesting to note that one can have interference between the
amplitude for lepton-quark scattering via photon exchange on one quark
with the amplitude for $Z^0$ exchange on another quark.  This implies a
distinctive parity-violating higher-twist contribution
$C^{PV}_{HT}(x)$ proportional to the product of electromagnetic and weak
quark charges
$\sum_{a \ne b} e^\gamma_a e^{Z^0}_b$.  Twist-four contributions of this
type have been in fact been modeled \cite{Castorina:1985uw} for
structure function moments.  However there is also the possibility of
high-$x$ enhancement.  In fact, the
$x$-dependence of $C^{PV}_{HT}(x)$ should be similar to the
parity-conserving contribution.

We can also use Bloom-Gilman duality to predict that the parity-violating
structure functions at large $x$ should average to the contributions of
the elastic and inelastic electroproduction channels when integrated over
similar ranges in $W^2$.  In fact, the parity-violating elastic form
factors can be predicted at large momentum transfer in
perturbative QCD.\cite{Brodsky:1981sx}  Such measurements will provide
very interesting tests of the applicability of PQCD to
exclusive processes.
Thus as emphasized by
Souder \cite{Souder:2000}, the detailed measurement of the left-right
asymmetry $A_{LR}$ in polarized elastic and inelastic electron-proton and
polarized electron nucleus scattering at large
$x_{bj}$ can be a powerful illuminator of quark-quark correlations and
fundamental QCD physics at the amplitude level.

\section{Higher Fock States and the Intrinsic Sea}

The main features of the heavy sea quark-pair contributions of the
higher particle number Fock state states of light hadrons can be
derived from perturbative QCD.  One can identify two contributions to the
heavy quark sea, the ``extrinsic'' contributions which correspond to
ordinary gluon splitting, and the ``intrinsic" sea which is
multi-connected via gluons to the valence quarks.  The leading $1/m_Q^2$
contributions to the intrinsic sea of the proton in the heavy quark
expansion are proton matrix elements of the operator~\cite{Franz:2000ee}
$\eta^\mu \eta^\nu G_{\alpha \mu} G_{\beta \nu} G^{\alpha \beta}$ which
in light-cone gauge $\eta^\mu A_\mu= A^+= 0$ corresponds to three or four
gluon exchange between the heavy-quark loop and the proton constituents
in the forward virtual Compton amplitude.  The intrinsic sea is thus
sensitive to the hadronic bound-state
structure.\cite{Brodsky:1981se,Brodsky:1980pb} The maximal contribution
of the intrinsic heavy quark occurs at $x_Q \simeq {m_{\perp Q}/ \sum_i
m_\perp}$ where
$m_\perp = \sqrt{m^2+k^2_\perp}$;
\ie\ at large $x_Q$, since this minimizes the invariant mass $\M^2_n$.
The
measurements of the charm structure function by the EMC experiment are
consistent with intrinsic charm at large $x$ in the nucleon with a
probability of order $0.6 \pm 0.3 \% $.\cite{Harris:1996jx} which is
consistent with recent estimates based on instanton
fluctuations.\cite{Franz:2000ee} Similarly, one can distinguish
intrinsic gluons which are associated with multi-quark interactions and
extrinsic gluon contributions associated with quark
substructure.\cite{Brodsky:1990db} One can also use this framework to
isolate the physics of the anomaly contribution to the Ellis-Jaffe sum
rule.\cite{Bass:1998rn} Thus neither gluons nor sea quarks are solely
generated by DGLAP evolution, and one cannot define a resolution scale
$Q_0$ where the sea or gluon degrees of freedom can be neglected.

It is usually assumed that a heavy quarkonium state such as the
$J/\psi$ always decays to light hadrons via the annihilation of its heavy quark
constituents to gluons.  However, as Karliner and I \cite{Brodsky:1997fj}
have shown, the transition $J/\psi \to \rho
\pi$ can also occur by the rearrangement of the $c \bar c$ from the $J/\psi$
into the $\ket{ q \bar q c \bar c}$ intrinsic charm Fock state of the $\rho$ or
$\pi$.  On the other hand, the overlap rearrangement integral in the
decay $\psi^\prime \to \rho \pi$ will be suppressed since the intrinsic
charm Fock state radial wavefunction of the light hadrons will evidently
not have nodes in its radial wavefunction.  This observation provides
a natural explanation of the long-standing puzzle~\cite{Brodsky:1987bb}
why the $J/\psi$ decays prominently to two-body pseudoscalar-vector final
states, breaking hadron helicity
conservation,~\cite{Brodsky:1981kj} whereas the
$\psi^\prime$ does not.

The higher Fock state of the proton $\ket{u u d s \bar s}$ should
resemble a $\ket{ K \Lambda}$ intermediate state, since this minimizes its
invariant mass $\M$.  In such a state, the
strange quark has a higher mean momentum fraction $x$ than the $\bar
s$.\cite{Warr,Signal,BMa}  Similarly, the helicity of the intrinsic
strange quark in this configuration will be anti-aligned with the
helicity of the nucleon.\cite{Warr,BMa}  This $Q \leftrightarrow \bar
Q$ asymmetry is a striking feature of the intrinsic heavy-quark sea.

\section{Other Applications of Light-Cone Quantization to QCD
 Phenomenology}

There are other phenomenological consequences of the light-cone Fock
expansion:

{\it Color Transparency.}
A crucial feature of the light-cone formalism is
the fact that the form of the
$\psi^{(\Lambda)}_{n/H}(x_i,
\vec k_{\perp i},\lambda_i)$ is invariant under longitudinal boosts; \ie,\ the
light-cone wavefunctions expressed in the relative coordinates $x_i$ and
$k_{\perp i}$ are independent of the total momentum
$P^+$,
$\vec P_\perp$ of the hadron.
The ensemble
\{$\psi_{n/H}$\} of such light-cone Fock
wavefunctions is a key concept for hadronic physics, providing a conceptual
basis for representing physical hadrons (and also nuclei) in terms of their
fundamental quark and gluon degrees of freedom.  Each Fock state interacts
distinctly; \eg, Fock states with small particle number and small impact
separation have small color dipole moments and can traverse a nucleus with
minimal interactions.  This is the basis for the predictions for ``color
transparency"  in hard quasi-exclusive reactions.\cite{BM}

{\it Diffractive vector meson photoproduction.} The
light-cone Fock wavefunction representation of hadronic amplitudes
allows a simple eikonal analysis of diffractive high energy processes, such as
$\gamma^*(Q^2) p \to \rho p$, in terms of the virtual photon and the vector
meson Fock state light-cone wavefunctions convoluted with the $g p \to g p$
near-forward matrix element.\cite{BGMFS}  One can easily show that only small
transverse size $b_\perp \sim 1/Q$ of the vector meson distribution
amplitude is involved.  The hadronic interactions are minimal, and thus the
$\gamma^*(Q^2) N \to
\rho N$ reaction can occur coherently throughout a nuclear target in reactions
without absorption or shadowing.  The $\gamma^* A \to V A$ process thus
is a laboratory for testing QCD color transparency.\cite{BM}

{\it Regge behavior of structure functions.} The light-cone wavefunctions
$\psi_{n/H}$ of a hadron are not independent of each other, but rather are
coupled via the equations of motion.  Antonuccio, Dalley and I
\cite{ABD} have used the constraint of finite ``mechanical'' kinetic energy to
derive ``ladder relations" which interrelate the light-cone wavefunctions of
states differing by one or two gluons.  We then use these relations to
derive the
Regge behavior of both the polarized and unpolarized structure functions at $x
\rightarrow 0$, extending Mueller's derivation of the BFKL hard
QCD pomeron from the properties of heavy quarkonium light-cone wavefunctions at
large $N_C$ QCD.\cite{Mueller}

{\it Structure functions at large $x_{bj}$.} The behavior of structure functions
where one quark has the entire momentum requires the knowledge of LC
wavefunctions
with $x \rightarrow 1$ for the struck quark and $x \rightarrow 0$ for the
spectators.  This is a highly off-shell configuration, and thus one can
rigorously derive quark-counting and helicity-retention rules for the
power-law behavior of
the polarized and unpolarized quark and gluon distributions in the $x
\rightarrow 1$ endpoint domain.

{\it DGLAP evolution at  $x \rightarrow 1$}
Usually one expects that structure functions are strongly suppressed at
large
$x$ because of the momentum lost by gluon radiation:  the predicted
change of the power law behavior at large $x$ is\cite{Gribov:1972ri}
\begin{equation}
\frac{F_2(x,Q^2)}{F_2(x,Q^2_0)} \eqx (1-x)^{\zeta(Q^2,Q^2_0)}
\label{eq:ai}
\end{equation}
where
\begin{equation}
\zeta(q^2,Q^2_0) = \frac{1}{4\pi} \int^{Q^2}_{Q^2_0}
\frac{d\ell^2}{\ell^2}\, \alpha_s(\ell^2) \ .
\label{eq:aj}
\end{equation}
Because of asymptotic freedom, this implies a $\log \log Q^2$ increase in
the effective power $\zeta(Q^2,Q^2_0)$.  However, this derivation assumes
that the struck quark is on its mass shell.  The off-shell effect is
profound, greatly reducing the PQCD
radiation.\cite{Brodsky:1979gy,Lepage:1982gd}
We can take into account the main effect of the struck quark virtuality by
modifying the propagator in Eq. (\ref{eq:aj}):
\begin{equation}
\zeta(Q^2,Q^2_0) = \frac{1}{4\pi} \int^{Q^2}_{Q^2_0}
\frac{d\ell^2}{\ell^2+|k^2_f|}\,
\alpha_s(\ell^2) .
\label{eq:ak}
\end{equation}
Thus at large $x$, there is effectively no DGLAP evolution until
$Q^2\gsim |k^2_f|$!  One can also see that DGLAP
evolution at large
$x$ at fixed $Q^2$ must be suppressed in order to have
duality at fixed
$W^2 = Q^2(1-x_{bj})/x_{bj}$ between the inclusive electroproduction and
exclusive resonance contributions.\cite{BL80}
Thus evolution
of structure functions is minimal in this domain because the struck quark
is highly virtual as $x\rightarrow 1$; \ie\ the starting point $Q^2_0$ for
evolution
cannot be held fixed, but must be larger than a scale of order
$(m^2 + k^2_\perp)/(1-x)$.\cite{LB,BrodskyLepage,Dmuller}

{\it Materialization of far-off-shell configurations.}
In a high energy hadronic collisions, the highly-virtual states of a hadron
can be
materialized into physical hadrons simply by the soft interaction of any of the
constituents.\cite{BHMT}  Thus a proton state with intrinsic charm $\ket{ u
u d \bar c c}$ can be materialized, producing a $J/\psi$ at large $x_F$,
by the
interaction of a light-quark in the target.  The production
occurs on the front-surface of a target nucleus, implying an $A^{2/3}$
$J/\psi$ production cross section at large $x_F,$  which is consistent
with experiment, such as Fermilab experiments E772 and E866.

{\it Comover phenomena.}
Light-cone wavefunctions describe not only the partons that interact in a
hard
subprocess but also the associated partons freed from the projectile.  The
projectile partons which are comoving (\ie, which have similar rapidity) with
final state quarks and gluons can interact strongly producing (a) leading
particle effects, such as those seen in open charm hadroproduction; (b)
suppression of quarkonium \cite{BrodskyMueller} in favor of open heavy hadron
production, as seen in the E772 experiment; (c) changes in color configurations
and selection rules in quarkonium hadroproduction, as has been emphasized by
Hoyer and Peigne.\cite{Hoyer:1998ha}   All of these effects violate the
usual ideas of factorization for inclusive reactions.  Further, more than
one parton from the
projectile can enter the hard subprocess, producing dynamical higher-twist
contributions, as seen for example in
Drell-Yan experiments.\cite{BrodskyBerger,Brandenburg}

{\it Jet hadronization in light-cone QCD.}
One of the goals of nonperturbative analysis in QCD is to compute jet
hadronization from first principles.  The DLCQ solutions provide a possible
method to accomplish this.  By inverting the DLCQ solutions, we can write the
``bare'' quark state of the free theory as
$\ket{q_0} = \sum \ket n \VEV{n\,|\,q_0}$
 where now $\{\ket n\}$ are the exact DLCQ eigenstates of
$H_{LC}$, and
$\VEV{n\,|\,q_0}$ are the DLCQ projections of the eigensolutions.  The expansion
in automatically infrared and ultraviolet regulated if we impose global cutoffs
on the DLCQ basis:
$\lambda^2 < \Delta\M^2_n < \Lambda^2
$
where $\Delta\M^2_n = \M^2_n-(\Sigma \M_i)^2$.  It would be
interesting to study jet hadronization at the amplitude level for
the existing DLCQ solutions to QCD (1+1) and collinear QCD.

{\it Hidden Color.}
The deuteron form factor at high $Q^2$ is sensitive to wavefunction
configurations where all six quarks overlap within an impact
separation $b_{\perp i} < {\cal O} (1/Q).$  The leading power-law
fall off predicted by QCD is $F_d(Q^2) = f(\alpha_s(Q^2))/(Q^2)^5$,
where, asymptotically, $f(\alpha_s(Q^2)) \propto
\alpha_s(Q^2)^{5+2\gamma}$.\cite{Brodsky:1976rz}  The derivation of the
evolution equation for the deuteron distribution amplitude and its
leading anomalous dimension $\gamma$ is given in  the references.\cite{bjl83}
In general, the six-quark wavefunction of a deuteron
is a mixture of five different color-singlet states.  The dominant
color configuration at large distances corresponds to the usual
proton-neutron bound state.  However at small impact space
separation, all five Fock color-singlet components eventually
acquire equal weight, \ie, the deuteron wavefunction evolves to
80\%\ ``hidden color.'' \cite{Brodsky:1983vf}
The relatively large normalization of the
deuteron form factor observed at large $Q^2$ hints at sizable
hidden-color contributions.\cite{Farrar:1991qi} Hidden color components
can also play a predominant role in the reaction $\gamma d \to J/\psi p n$
at threshold if it is dominated by the multi-fusion process $\gamma g g
\to J/\psi$.

{\it Nuclear Structure Functions at $1 < x_{bj} < A$, beyond the
kinematic domain accessible on a single nucleon target.}  The nuclear
light-cone momentum must be transferred to a single quark, requiring
quark-quark correlations between quarks of different nucleons in a
compact, far-off-shell regime.  Also, as noted above, the nuclear
wavefunction contains hidden-color components distinct from a convolution
of separate color-singlet nucleon wavefunctions.

{\it Spin-Spin Correlations in Nucleon-Nucleon
Scattering and the Charm \hfill\break Threshold.}
One of the most striking anomalies in elastic proton-proton
scattering is the large spin correlation $A_{NN}$ observed at large
angles.\cite{krisch92}  At $\sqrt s \simeq 5 $ GeV, the rate for
scattering with incident proton spins parallel and normal to the
scattering plane is four times larger than that for scattering with
anti-parallel polarization.  This strong polarization correlation can
be attributed to the onset of charm production in the intermediate
state at this energy.\cite{Brodsky:1988xw}  The intermediate state $\vert u
u d u u d c \bar c \rangle$ has odd intrinsic parity and couples to
the $J=S=1$ initial state, thus strongly enhancing scattering when
the incident projectile and target protons have their spins parallel
and normal to the scattering plane.  The charm threshold can also
explain the anomalous change in color transparency observed at the
same energy in quasi-elastic $ p p$ scattering.  A crucial test is
the observation of open charm production near threshold with a
cross
section of order of $1 \mu$b.

\section{Conclusions}

In these lectures I have
discussed how the universal, process-independent and
frame-independent light-cone Fock-state wavefunctions
encode the properties of a hadron in terms of its fundamental quark and
gluon degrees of freedom.  Given the proton's light-cone wavefunctions,
one can compute not only the moments of the quark and gluon distributions
measured in deep inelastic lepton-proton scattering, but also the
multi-parton correlations which control the distribution of particles in
the proton fragmentation region and dynamical higher twist effects.
Light-cone wavefunctions also provide a systematic framework for
evaluating exclusive hadronic matrix elements, including time-like heavy
hadron decay amplitudes and form factors.  The formalism also provides a
physical factorization scheme for separating hard and soft contributions
in both exclusive and inclusive hard processes.
I have discussed a number of applications of light-cone Fock
representation of QCD, including semileptonic $B$ decays,
deeply virtual Compton scattering, and dynamical higher twist effects in
inclusive reactions.  The relation of the
intrinsic sea to the light-cone wavefunctions is discussed.  The physics
of light-cone wavefunctions is illustrated for the quantum fluctuations
of an electron.
A new type of jet
production reaction, ``self-resolving diffractive interactions" can
provide direct information on the light-cone wavefunctions of hadrons in
terms of their QCD degrees of freedom, as well as the composition of
nuclei in terms of their nucleon and mesonic degrees of freedom.
I have also reviewed the strong progress that has been made in computing
light-cone wavefunctions directly from the QCD light-cone Hamiltonian.
Even without full non-perturbative solutions of QCD, one can envision a
program to construct the light-cone wavefunctions using measured moments
constraints from QCD sum rules, lattice gauge theory,  hard exclusive and
inclusive processes.  One is guided by theoretical constraints from
perturbation theory which dictates the asymptotic form of the
wavefunctions at large invariant mass,
$x \to 1$, and high
$k_\perp$.  One can also use constraints
from ladder relations which connect Fock states of
different particle number; perturbatively-motivated numerator spin
structures; conformal symmetry, guidance from toy models
such as ``reduced"
$QCD(1+1)$; and the correspondence to Abelian theory
for
$N_C\to 0$, and the many-body
Schr\"odinger theory in the nonrelativistic domain.

\section*{Acknowledgments}
Work supported by the Department of Energy
under contract number DE-AC03-76SF00515.
I wish to thank Fernando Navarra and Martina Nielson
for their kind hospitality in Brazil.  Much of this work is based on
collaborations, particularly with  Markus Diehl, Paul Hoyer,
Dae Sung Hwang,  Peter Lepage, Bo-Qiang Ma,  Hans Christian Pauli, Johan
Rathsman,  Ivan Schmidt, and Prem Srivastava.

\end{document}